\documentclass[aps,twocolumn,showpacs,pre]{revtex4-1}
\usepackage{graphicx,graphics}
\usepackage{amsmath,amssymb,amsfonts}
\usepackage{bbm,bbold}
\usepackage{multirow}
\usepackage{hyperref}
\usepackage{graphicx}
\usepackage{color}

\begin{document}
\title{Diffusive transport on networks with stochastic resetting to multiple nodes }
\author{Fernanda H. Gonz\'alez$^1$}
\author{Alejandro P. Riascos$^1$}
\author{Denis Boyer$^1$}
\affiliation{${}^1$Instituto de F\'isica,Universidad Nacional Aut\'onoma de M\'exico,\\
Apartado Postal 20-364, 01000 Ciudad de M\'exico, M\'exico}
\date{\today}
\begin{abstract}
We study the diffusive transport of Markovian random walks on arbitrary networks with stochastic resetting to multiple nodes. We deduce analytical expressions for the stationary occupation probability and for the mean and global first passage times. This general approach allows us to characterize the effect of resetting on the capacity of random walk strategies to reach a particular target or to explore the network. Our formalism holds for ergodic random walks and can be implemented from the spectral properties of the random walk without resetting, providing a tool to analyze the efficiency of search strategies with resetting to multiple nodes. We apply the methods developed here to the dynamics with two reset nodes and derive analytical results for normal random walks and L\'evy flights on rings. We also explore the effect of resetting to multiple nodes on a comb graph, L\'evy flights that visit specific locations in a continuous space, and the Google random walk strategy on regular networks.
\end{abstract}


\maketitle
\section{Introduction}
Diffusive transport and random walk strategies have been implemented in diverse fields as processes that are able to efficiently reach hidden targets or to simply explore a particular region of space. Examples include animal foraging \cite{ViswaBook2011}, the activity of urban transportation systems \cite{LoaizaMonsalvePlosOne2019,RiascosMateosSciRep2020}, protein searching for specific binding sites on the DNA \cite{Coppey2004}, searching and ranking databases \cite{LeskovecBook2014,BlanchardBook2011}, among many others. In this context, there has been in recent years an increasing interest in search processes with resetting or restart. When a stochastic process is occasionally reset, {\it i.e.}, interrupted and restarted from the initial state, its dynamics is strongly altered. Interestingly, the average time needed to reach a given target state for the first time can often be minimized with respect to the resetting rate \cite{evans2011diffusion,Evans2011JPhysA,reuveni2016optimal,EvansReview2019}.
Different types of resetting protocols have been considered \cite{pal2016diffusion,nagar2016diffusion,Bhat2016JStat,chechkin2018random} on a variety of underlying processes, such as Brownian motion \cite{evans2011diffusion,Evans2011JPhysA,MajumdarPRE2015}, processes with a drift \cite{montero2013monotonic,ray2019peclet} or models of anomalous diffusion \cite{Kusmierz2014PRL,kusmierz2015optimal,kusmierz2019subdiffusive,maso2019transport}.
\\[2mm]
In addition, a huge variety of phenomena can be described in terms of dynamical processes on networks \cite{NewmanBook,barabasi2016book}. The interplay between the topology of a network and the dynamical processes taking place on it is the key to understanding many complex systems \cite{NewmanBook,VespiBook,VanMieghem2011}. In particular, random walk strategies that allow transitions between nearest-neighbor nodes on a network are relevant to many problems and constitute the natural framework to study diffusive transport \cite{VespiBook,Hughes,Lovasz1996,MulkenPR502}. Network exploration by random walks is now better understood \cite{NohRieger2004,Tejedor2009PRE,MasudaPhysRep2017}, including non-local strategies with long-range hops between distant nodes \cite{RiascosMateos2012,RiascosMateosFD2014,Weng2015,Guo2016,Michelitsch2017PhysA,deNigris2017,Estrada2017Multihopper}, and the collective activity of simultaneous random walkers \cite{WengPRE2017,WengChaos2018,AgliariPRE2016,AgliariPRE2019,Riascos_PREMultiple2021}. Random walks on networks under the influence of resetting have been relatively little explored \cite{Avrachenkov2014,Avrachenkov2018,Touchette_PRE2018, ResetNetworks_PRE2020,christophorov2020peculiarities,Wald_PRE2021,bonomo2021first}. A couple of recent studies have established relationships between the random walk dynamics with resetting to one node and the spectral representation of the transition matrix that defines the random walk without resetting \cite{Touchette_PRE2018,ResetNetworks_PRE2020}. These results highlight that processes under resetting are promising strategies for exploring different network topologies \cite{ResetNetworks_PRE2020}.
\\[2mm]
\begin{figure}[!b]
	\begin{center}
		\includegraphics*[width=0.4\textwidth]{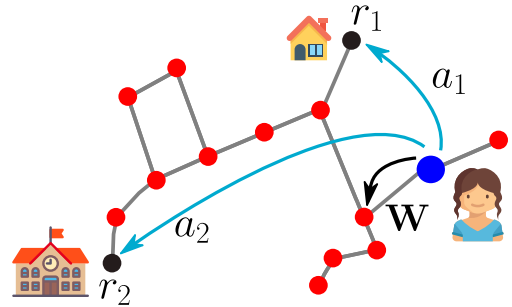}
	\end{center}
	\vspace{-5mm}
	\caption{\label{Fig_1} A random walker under resetting is illustrated as an agent visiting different points of interest in a network, with restart from specific locations. From any node in the network, transitions to the nodes $r_1$ (home) and $r_2$ (work) occur with probabilities $a_1$ and $a_2$, respectively. Otherwise, the visits to other sites are described by a transition probability matrix $\mathbf{W}$. }
\end{figure}
In this paper, we extend the spectral methods developed in Ref. \cite{ResetNetworks_PRE2020} to the analysis of random walk strategies with resetting to multiple nodes in the network. Figure \ref{Fig_1} exemplifies this process with an agent that visits different nodes (say, points of interest in a city) with frequent returns to two specific sites of major importance. Dynamical processes that consider a set of nodes to which stochastic resetting can occur find applications in different contexts; for example, the modeling of routines in human mobility \cite{Pappalardo2015,Schneider2013Interface}, problems of label propagation in machine learning algorithms \cite{Bautista2019}, or the Google strategy, which can be interpreted as a random walker with uniform resetting probability to all the nodes of the network \cite{Brin1998,ShepelyanskyRevModPhys2015}. Some related problems were addressed in Ref. \cite{Evans2011JPhysA} in continuous spaces, namely, a Brownian motion on the line with resetting to a random position drawn from a given resetting distribution. Recently, Besga {\it et al.} unveiled that the mean first passage times of one-dimensional Brownian particles with resetting to a random position with Gaussian distribution could exhibit behaviors markedly different from single-point resetting, depending on the width of the Gaussian distribution \cite{Besga_PhysRevRes_2020}. Despite these studies, resetting processes to multiple points remain little understood, especially in the context of networks.
\\[2mm]
The paper is organized as follows. We begin with a summary of the main results relative to the analysis of ergodic random walks with resetting to one node. The detailed extension of these results to two resetting nodes is further developed for general random walks. We deduce analytical expressions for the stationary distribution (or occupation probability) and for the mean first passage times. We explore several examples, such as two cases on the ring topology: normal random walks with transitions to nearest-neighbor sites and L\'evy flights where transitions between distant nodes are possible. The latter non-local dynamics are generated by considering the fractional Laplacian of the network. Then, we extend our analysis to $\mathcal{M}$ resetting nodes. We further apply our methods to study the overall effect of resetting to multiple nodes on a comb graph, a random walker that visits specific locations in a continuous space, and on a Google random walker on regular networks. The methods introduced here provide a general framework to obtain analytically the eigenvalues and eigenvectors of operators with reset to multiple nodes, and can be implemented to study different dynamical processes with restart.
\section{Random walks with resetting to one node}
\label{RW_reset1}
Let us consider an ergodic random walk on an arbitrary connected network with $N$ nodes $i=1,\ldots ,N$. We study the random walker in discrete time $t=0,1,2,\ldots$ starting at $t=0$ from a node $i$. The walker performs two types of steps: with probability $1-\gamma$, a random jump from the node currently occupied to a different node of the network, or, with probability $\gamma$, a resetting to a fixed node $r$. Without resetting ($\gamma=0$), the probability to hop to $m$ from $l$ is denoted as $w_{l\to m}$, and we assume that the random walk is ergodic and described by the transition matrix $\mathbf{W}$ with elements $w_{l\to m}$ for $l,m=1,\ldots,N$. The transition matrix is general in the sense that it can be local, {\it i.e.}, with transitions only between connected nodes (that we will denote here as ``nearest neighbors''), or non-local, including jumps between distant nodes, that are not directly connected to each other. 
\\[2mm]
The occupation probability of the process under resetting follows the master equation \cite{ResetNetworks_PRE2020}
\begin{equation}
 \label{mastereq1}
 P_{ij}(t+1;r,\gamma) = (1-\gamma)\sum_{l=1}^N P_{il}(t;r,\gamma)w_{l\to j}+\gamma\delta_{rj},
\end{equation}
here $P_{ij}(t;r,\gamma)$ denotes the probability to find the walker at $j$ at time $t$, given the initial position $i$, resetting node $r$ and resetting probability $\gamma$ ($\delta_{rj}$ denotes the Kronecker delta).
The first term on the right-hand side of Eq. (\ref{mastereq1}) represents hops associated to the transition probabilities  $\mathbf{W}$ and the second term describes resetting to $r$. With the introduction of the transition probability matrix $\mathbf{\Pi}(r;\gamma)$ with elements $
\pi_{l \to m}(r;\gamma)\equiv (1-\gamma) w_{l\to m}+\gamma\,\delta_{rm}$, Eq. (\ref{mastereq1}) takes the simpler form \cite{ResetNetworks_PRE2020}
\begin{equation}\label{mastermarkov}
P_{ij}(t+1;r,\gamma) = \sum_{l=1}^N  P_{il} (t;r,\gamma) \pi_{l\to j}(r;\gamma),
\end{equation}
where $\sum_{m=1}^N \pi_{l \to m}(r;\gamma)=1$. The matrix
$\mathbf{\Pi}(r;\gamma)$ completely entails the process with resetting, which is able to reach all the nodes of the network if the resetting probability $\gamma$ is $<1$. The matrices $\mathbf{W}$ and $\mathbf{\Pi}(r;\gamma)$ are stochastic matrices: knowing their eigenvalues and eigenvectors allows the calculation of the occupation probability at any time, including the stationary distribution at $t=\infty$, as well as the mean first passage time to any node. The eigenvalues and eigenvectors of $\mathbf{\Pi}(r;\gamma)$  are related to those of $\mathbf{W}$, which is recovered in the limit $\gamma=0$ \cite{ResetNetworks_PRE2020}.
\\[2mm]
In the following we use Dirac's notation for eigenvectors. We denote the eigenvalues of the matrix $\mathbf{W}$ as $\lambda_l$ (where $\lambda_1=1$), and its right and left eigenvectors as $\left|\phi_l\right\rangle$ and $\left\langle\bar{\phi}_l\right|$, respectively, for $l=1,2,\ldots,N$. These eigenvectors form an orthonormal base and satisfy the relations 
\begin{equation}\label{phi_vec_conds}
\left\langle\bar{\phi}_l|\phi_m\right\rangle=\delta_{lm}, \qquad \sum_{l=1}^N \left|\phi_l\right\rangle \left\langle\bar{\phi}_l\right|=\mathbb{I},
\end{equation}
being $\mathbb{I}$ the $N\times N$ identity matrix. Similarly, the eigenvalues of $\mathbf{\Pi}(r;\gamma)$ are denoted as $\zeta_l(r;\gamma)$ and its eigenvectors as $\left|\psi_l(r;\gamma)\right\rangle$ and $\left\langle\bar{\psi}_l(r;\gamma)\right|$.
\\[2mm]
The connection between the eigenvalues $\lambda_l$ and $\zeta_l(r;\gamma)$ is obtained from the relation 
\begin{equation}\label{Def_MatPi_R1}
\mathbf{\Pi}(r;\gamma)=(1-\gamma)\mathbf{W}+\gamma \mathbf{\Theta}(r),
\end{equation}
where the elements of the matrix $\mathbf{\Theta}(r)$ are $\Theta_{lm}(r)=\delta_{mr}$. Namely, $\mathbf{\Theta}(r)$ has entries $1$ in the $r$th column and null entries everywhere else, therefore (see Ref. \cite{ResetNetworks_PRE2020} for details),
\begin{equation}\label{eigvals_zeta}
\zeta_l(r;\gamma)=
\begin{cases}
1 \qquad &\mathrm{for}\qquad l=1,\\
(1-\gamma)\lambda_l \qquad &\mathrm{for}\qquad l=2,3,\ldots, N.
\end{cases}
\end{equation}
This result reveals that the eigenvalues are independent of the choice of the resetting node $r$. On the other hand, the left eigenvectors of $\mathbf{\Pi}(r;\gamma)$ are given by  \cite{ResetNetworks_PRE2020}
\begin{equation}\label{psil1}
\left\langle\bar{\psi}_1(r;\gamma)\right|=\left\langle\bar{\phi}_1\right|
+\sum_{m=2}^N\frac{\gamma}{1-(1-\gamma)\lambda_m}\frac{\left\langle r|\phi_m\right\rangle}{\left\langle r|\phi_1\right\rangle}\left\langle\bar{\phi}_m\right|
\end{equation}
whereas $\left\langle\bar{\psi}_l(r;\gamma)\right|=\left\langle\bar{\phi}_l\right|$ for $l=2,\ldots,N$. Similarly, the right eigenvectors are given by:
$\left|\psi_1(r;\gamma)\right\rangle=\left|\phi_1\right\rangle$ and 
\begin{equation}\label{psirl_reset}
\left|\psi_l(r;\gamma)\right\rangle=
\left|\phi_l\right\rangle-\frac{\gamma}{1-(1-\gamma)\lambda_l}\frac{\left\langle r|\phi_l\right\rangle }{\left\langle r|\phi_1\right\rangle}\left|\phi_1\right\rangle,
\end{equation}
for $l=2,\ldots,N$, where $|r\rangle$ denotes the vector with all its components equal to 0 except the $r$-th one, which is equal to 1\cite{ResetNetworks_PRE2020}.
\\[2mm]
With the left and right eigenvectors at hand, one can use the spectral representation
\begin{equation}\label{Pi_prob_powert}
\mathbf{\Pi}(r;\gamma)=\sum_{l=1}^N\zeta_l(r;\gamma)\left|\psi_l(r;\gamma)\right\rangle\left\langle\bar{\psi}_l(r;\gamma)\right|.
\end{equation}
In this notation, the occupation probability of the process described by Eq. (\ref{mastermarkov}) is  \cite{ResetNetworks_PRE2020}
\begin{align}\nonumber
&P_{ij}(t;r,\gamma)=P_j^\infty(r;\gamma)\\
&+\sum_{l=2}^N(1-\gamma)^t\lambda_l^t\left[\left\langle i|\phi_l\right\rangle \left\langle\bar{\phi}_l|j\right\rangle-\gamma\frac{\left\langle r|\phi_l\right\rangle \left\langle\bar{\phi}_l|j\right\rangle}{1-(1-\gamma)\lambda_l} \right], \label{Pijspect}
\end{align}
where $|i\rangle$ and $|j\rangle$ are defined similarly to $|r\rangle$.
The first term of the right-hand side in Eq. (\ref{Pijspect}) defines the long time stationary distribution $P_j^\infty(r;\gamma)=\left\langle i\left|\psi_1(r;\gamma)\right\rangle \left\langle\bar{\psi}_1(r;\gamma)\right|j\right\rangle$. By using  Eq. (\ref{psil1}) and $\left|\psi_1(r;\gamma)\right\rangle=\left|\phi_1\right\rangle$, one obtains \cite{ResetNetworks_PRE2020}
\begin{equation}\label{Pinfvectors_1}
P_j^\infty(r;\gamma)=P_i^\infty(0)+\gamma\sum_{l=2}^N\frac{\left\langle r|\phi_l\right\rangle \left\langle\bar{\phi}_l|j\right\rangle}{1-(1-\gamma)\lambda_l},
\end{equation}
where we have used the identity $P_i^\infty(0)=\left\langle i|\phi_1\right\rangle \left\langle\bar{\phi}_1|j\right\rangle$ for the equilibrium distribution of the random walk without resetting
\cite{NohRieger2004,MasudaPhysRep2017}.
\\[2mm]
As for the asymptotic distribution in Eq. (\ref{Pinfvectors_1}), the occupation probability at finite time $P_{ij}(t;r,\gamma)$ is expressed in terms of the eigenvalues and eigenvectors of $\mathbf{W}$. By using the well-known convolution for Markov processes between $P_{ij}$ at time $t$ and the first passage time distribution \cite{NohRieger2004,Hughes} (see Appendix \ref{App_DeductionTij} for details), we deduce the exact expression of the mean first passage time (MFPT) at node $j$ when starting from $i$, $\langle T_{ij}(r;\gamma)\rangle$
\begin{equation}\label{Tij_R}
\langle T_{ij}(r;\gamma)\rangle=\frac{\mathcal{R}_{jj}^{(0)}(r;\gamma)-\mathcal{R}_{ij}^{(0)}(r;\gamma)+\delta _{ij}}{P_j^{\infty}(r;\gamma)},
\end{equation}
with the moments  $\mathcal{R}^{(n)}_{ij}(r;\gamma)$ defined as
\begin{equation}\label{Rmoments_def}
\mathcal{R}^{(n)}_{ij}(r;\gamma)\equiv \sum_{t=0}^{\infty} t^n ~ \{P_{ij}(t;r,\gamma)-P_j^\infty(r;\gamma)\}.
\end{equation}
Hence, combining Eqs. (\ref{Pijspect})-(\ref{Rmoments_def}), one obtains the MFPT for a random walker that starts at $i$ and reaches for the first time $j$, subject to stochastic resetting to the node $r$ \cite{ResetNetworks_PRE2020}
\begin{multline}\label{MFPT_resetSM}
\left\langle T_{ij}(r;\gamma)\right\rangle=\frac{\delta_{ij}}{P_j^\infty(r;\gamma)}\\+
\frac{1}{P_j^\infty(r;\gamma)}\sum_{\ell=2}^N\frac{
\left\langle j|\phi_\ell\right\rangle \left\langle\bar{\phi}_\ell|j\right\rangle-\left\langle i|\phi_\ell\right\rangle \left\langle\bar{\phi}_\ell|j\right\rangle
}{1-(1-\gamma)\lambda_\ell}.
\end{multline}
\section{Random walks with resetting to two nodes}
\label{RW_reset2}
The results of Section \ref{RW_reset1} describe the effect of random walk resetting to one specific node in terms of the eigenvalues and eigenvectors of the transition matrix $\mathbf{W}$. In this section, we generalize this formalism to consider resetting to two nodes. We present the general results for ergodic random walks and analyze local and non-local random walks on rings.
\subsection{General approach}
\label{TwoNodes_General}
Now, let us explore the dynamics with reset to $\mathcal{M}=2$ nodes $r_1$, $r_2$ with probabilities $a_1$ and $a_2$. The transition matrix of this process is
\begin{equation}\label{matPi_2reset}
\mathbf{\Pi}(r_1,r_2;a_1,a_2)=a_0\mathbf{W}+a_1\mathbf{\Theta}(r_1)+a_2\mathbf{\Theta}(r_2),
\end{equation}
with $a_0=1-a_1-a_2$. Equation (\ref{matPi_2reset}) can be reorganized as follows
\begin{multline}
\mathbf{\Pi}(r_1,r_2;a_1,a_2)=\left(a_0+a_1\right)\times\\\Big[\frac{a_0}{a_0+a_1}\mathbf{W}
+\frac{a_1}{a_0+a_1}\mathbf{\Theta}(r_1)\Big]+a_2\mathbf{\Theta}(r_2),
\end{multline}
or, by defining 
\begin{equation}
\gamma_1\equiv \frac{a_1}{a_0+a_1},
\end{equation}
we have
\begin{align}\nonumber
&\mathbf{\Pi}(r_1,r_2;a_1,a_2)\\\nonumber
&=\left(a_0+a_1\right)\left[(1-\gamma_1)\mathbf{W}+\gamma_1\mathbf{\Theta}(r_1)\right]+a_2\mathbf{\Theta}(r_2)\\ \nonumber
&=\left(a_0+a_1\right)\mathbf{\Pi}(r_1;\gamma_1)+a_2\mathbf{\Theta}(r_2)\\ \nonumber
&=\left(a_0+a_1+a_2\right)\times\\
&\hspace{5mm}\Big[\frac{a_0+a_1}{a_0+a_1+a_2}\mathbf{\Pi}(r_1;\gamma_1)+\frac{a_2}{a_0+a_1+a_2}\mathbf{\Theta}(r_2)\Big],
\end{align}
where we used the matrix $\mathbf{\Pi}(r_1;\gamma_1)$ given by Eq. (\ref{Def_MatPi_R1}) for the dynamics with reset to the node $r_1$ with probability $\gamma_1$. Using the fact that $a_0+a_1+a_2=1$ and defining $\gamma_2\equiv a_2$, $\mathbf{\Pi}(r_1,r_2;a_1,a_2)$ can be expressed as
\begin{equation}\label{matPi_iterative2}
\mathbf{\Pi}(r_1,r_2;a_1,a_2)=(1-\gamma_2)\mathbf{\Pi}(r_1;\gamma_1)+\gamma_2\mathbf{\Theta}(r_2).
\end{equation}
Since we know all the eigenvalues and eigenvectors of $\mathbf{\Pi}(r_1;\gamma_1)$, we can apply the analysis of the case with one reset node to write $\mathbf{\Pi}(r_1,r_2;a_1,a_2)$ in terms of  $\mathbf{\Pi}(r_1;\gamma_1)$, and then express the results in terms of the eigenvalues and eigenvectors of the original $\mathbf{W}$.
\\[2mm]
The analysis for the eigenvalues leads to $\zeta_1(r_1,r_2;\gamma_1,\gamma_2)=1$ and
\begin{equation}
\zeta_l(r_1,r_2;\gamma_1,\gamma_2)=
(1-\gamma_2)(1-\gamma_1)\lambda_l,\,\,
\end{equation}
for $l=2,\ldots, N$. In a similar way, we represent the first left eigenvector of $\mathbf{\Pi}(r_{1},r_{2};a_1,a_2)$ in terms of the eigenvectors of  $\mathbf{\Pi}(r_{1};\gamma_{1})$
\begin{multline} \label{eq:left_1_complete}
\left\langle\bar{\psi}_1(r_1,r_2;\gamma_1,\gamma_2)\right|=\left\langle\bar{\psi}_1(r_1;\gamma_1)\right|\\
+\sum_{m=2}^N\frac{\gamma_2}{1-(1-\gamma_2)\zeta_m(r_1;\gamma_1)}\times
\\\frac{\left\langle r_2|\psi_m(r_1;\gamma_1)\right\rangle}{\left\langle r_2|\psi_1(r_1;\gamma_1)\right\rangle}\left\langle\bar{\psi}_m(r_1;\gamma_1)\right|.
\end{multline}
Therefore, considering the results for the reset to one node, we get
\begin{multline} \label{eq:left_1_gammas}
\left\langle\bar{\psi}_1(r_1,r_2;\gamma_1,\gamma_2)\right|=\left\langle\bar{\phi}_1\right|\\
+\sum_{m=2}^N\frac{\gamma_1}{1-(1-\gamma_1)\lambda_m}\frac{\left\langle r_1|\phi_m\right\rangle}{\left\langle r_1|\phi_1\right\rangle}\left\langle\bar{\phi}_m\right|\\
+\sum_{m=2}^N\frac{\gamma_2}{1-(1-\gamma_2)(1-\gamma_1)\lambda_m}\bigg[\frac{\left\langle r_2|\phi_m\right\rangle }{\left\langle r_2|\phi_1\right\rangle}\\
-\frac{\gamma_1}{1-(1-\gamma_1)\lambda_m}\frac{\left\langle r_1|\phi_m\right\rangle }{\left\langle r_1|\phi_1\right\rangle}\bigg]\left\langle\bar{\phi}_m\right|
\end{multline}
or, after rearranging the terms,
\begin{multline}\label{EigenL1_M2}
\left\langle\bar{\psi}_1(r_1,r_2;\gamma_1,\gamma_2)\right|=\left\langle\bar{\phi}_1\right|\\
+\sum_{m=2}^N\frac{\gamma_1}{1-(1-\gamma_1)\lambda_m}\frac{\left\langle r_1|\phi_m\right\rangle}{\left\langle r_1|\phi_1\right\rangle}\times \\
\bigg[1-\frac{\gamma_2}{1-(1-\gamma_2)(1-\gamma_1)\lambda_m}\bigg]\left\langle\bar{\phi}_m\right|\\
+\sum_{m=2}^N\frac{\gamma_2}{1-(1-\gamma_2)(1-\gamma_1)\lambda_m}\frac{\left\langle r_2|\phi_m\right\rangle }{\left\langle r_2|\phi_1\right\rangle}\left\langle\bar{\phi}_m\right|.
\end{multline}
In order to have a more compact notation, we define the coefficients
\begin{equation}\label{eq:nu}
\nu_{m}\equiv\frac{\gamma_2}{1-(1-\gamma_2)(1-\gamma_1)\lambda_m}
\end{equation}
and
\begin{equation*}
\kappa_{m}\equiv\frac{\gamma_1}{1-(1-\gamma_1)\lambda_m}\left(1-\frac{\gamma_2}{1-(1-\gamma_2)(1-\gamma_1)\lambda_m}\right)\\
\end{equation*}
which are also related through
\begin{equation}
\label{eq:kappa}
\kappa_{m}=\frac{\gamma_1}{1-(1-\gamma_1)\lambda_m}(1-\nu_{m}).
\end{equation}
It is important to bear in mind that $\nu_{m}$ and $\kappa_{m}$, depend on $\gamma_{1}$, $\gamma_{2}$ and the eigenvalues $\lambda_{m}$. Hence, the first left eigenvector in Eq. (\ref{EigenL1_M2}) 
is given by
\begin{multline} \label{eq:left_1_kappa_nu}
\left\langle\bar{\psi}_1(r_1,r_2;\gamma_1,\gamma_2)\right|=\left\langle\bar{\phi}_1\right|\\
+\sum_{m=2}^N\left(\kappa_{m}\frac{\left\langle r_1|\phi_m\right\rangle}{\left\langle r_1|\phi_1\right\rangle}
+\nu_{m}\frac{\left\langle r_2|\phi_m\right\rangle }{\left\langle r_2|\phi_1\right\rangle}\right)\left\langle\bar{\phi}_m\right|.
\end{multline}
On the other hand, for $l=2,3,\ldots, N$, the left eigenvectors of $\mathbf{\Pi}(r_{1},r_{2};;a_1,a_2)$ are given directly by the left eigenvectors of $\mathbf{W}$, therefore
\begin{equation} \label{eq:left_l}
\left\langle\bar{\psi}_l(r_1,r_2;\gamma_1,\gamma_2)\right|=\left\langle\bar{\phi}_l(r_1;\gamma_1)\right|=\left\langle\bar{\phi}_l\right|.
\end{equation}
Following a similar procedure for the first right eigenvector of $\mathbf{\Pi}(r_1,r_2;a_1,a_2)$, we obtain
\begin{equation} \label{eq:right_1}
    |\psi_{1}(r_{1},r_{2};\gamma_{1},\gamma_{2})\rangle = |\psi_{1}(r_{1};\gamma_{1})\rangle = |\phi_{1}\rangle.
\end{equation}
If $l=2,\ldots,N$ we apply Eq. (\ref{psirl_reset}) to get
\begin{multline*}
    |\psi_{l}(r_{1},r_{2};\gamma_{1},\gamma_{2})\rangle = |\psi_{l}(r_{1};\gamma_{1})\rangle\\
    -\frac{\gamma_{2}}{1-(1-\gamma_{2})(1-\gamma_{1})\lambda_{l}}\frac{\langle r_{2}|\psi_{l}(r_{1};\gamma_{1})\rangle}{\langle r_{2}|\psi_{1}(r_{1};\gamma_{1}) \rangle}|\psi_{1}(r_{1};\gamma_{1})\rangle,
\end{multline*}
and, by substituting the corresponding eigenvectors of $\mathbf{\Pi}(r_1;\gamma_1)$,
\begin{multline}
    |\psi_{l}(r_{1},r_{2};\gamma_{1},\gamma_{2})\rangle = |\phi_{l}\rangle- \bigg[\frac{\gamma_{1}}{1-(1-\gamma_{1})\lambda_{l}}\frac{\langle r_{1}|\phi_{l}\rangle}{\langle r_{1}|\phi_{1}\rangle}\times\\
    \bigg(1-\frac{\gamma_{2}}{1-(1-\gamma_{2})(1-\gamma_{1})\lambda_{l}}\bigg)\\
    +\frac{\gamma_{2}}{1-(1-\gamma_{2})(1-\gamma_{1})\lambda_{l}}\frac{\langle r_{2}|\phi_{l}\rangle}{\langle r_{2}|\phi_{1}\rangle}\bigg]|\phi_{1}\rangle.
\end{multline}
In this manner, $|\psi_{l}(r_{1},r_{2};\gamma_{1},\gamma_{2})\rangle$ can be expressed in terms of $\kappa_{m}$ and $\nu_{m}$, for $l=2,\ldots,N$,
\begin{multline} \label{eq:right_l_kappa_nu}
    |\psi_{l}(r_{1},r_{2};\gamma_{1},\gamma_{2})\rangle = |\phi_{l}\rangle\\
- \bigg(\kappa_{l}\frac{\langle r_{1}|\phi_{l}\rangle}{\langle r_{1}|\phi_{1}\rangle}
    +\nu_{l}\frac{\langle r_{2}|\phi_{l}\rangle}{\langle r_{2}|\phi_{1}\rangle}\bigg)|\phi_{1}\rangle.
\end{multline}
Given these eigenvectors, we deduce the stationary distribution
\begin{align*}
&P_j^\infty(r_1,r_2;\gamma_1,\gamma_2)\\
&=\left\langle i\left|\psi_1(r_1,r_2;\gamma_1,\gamma_2)\right\rangle \left\langle\bar{\psi}_1(r_1,r_2;\gamma_1,\gamma_2)\right|j\right\rangle\\
&=P_i^\infty(0)+ \sum_{m=2}^N\frac{\gamma_1}{1-(1-\gamma_1)\lambda_m}\frac{\left\langle r_1|\phi_m\right\rangle}{\left\langle r_1|\phi_1\right\rangle}\times\\
&\hspace{7mm}\bigg[1-\frac{\gamma_2}{1-(1-\gamma_2)(1-\gamma_1)\lambda_m}\bigg]\langle i|\phi_{1}\rangle\left\langle\bar{\phi}_m\right|j\rangle\\
&\hspace{3mm}+\sum_{m=2}^N\frac{\gamma_2}{1-(1-\gamma_2)(1-\gamma_1)\lambda_m}\frac{\left\langle r_2|\phi_m\right\rangle }{\left\langle r_2|\phi_1\right\rangle}\langle i|\phi_{1}\rangle\left\langle\bar{\phi}_m\right|j\rangle.
\end{align*}
From this result, it is clear that if either $\gamma_{1}$ or $\gamma_{2}$ are zero, we recover the stationary distribution for one resetting node given by Eq. (\ref{Pinfvectors_1}). $P_j^\infty(r_1,r_2;\gamma_1,\gamma_2)$ can be further simplified by noting that $|\phi_1\rangle$, associated to $\lambda_1=1$, has constant entries due to the normalization condition $\sum_{j}w_{i\to j}=1$. Therefore $\langle l |\phi_{1}\rangle=\mathrm{constant}$ for all $l$ and using the coefficients in Eqs. \eqref{eq:nu} and \eqref{eq:kappa}, we get
\begin{multline}
\label{eq:Pj_infty_final}
P_j^\infty(r_1,r_2;\gamma_1,\gamma_2)=P_i^\infty(0)\\+ \sum_{m=2}^N\left(\kappa_{m}\left\langle r_1|\phi_m\right\rangle+\nu_{m}\left\langle r_2|\phi_m\right\rangle\right)\left\langle\bar{\phi}_m\right|j\rangle.
\end{multline}
The general relation for the occupation probability at finite time $t$,
\begin{align}\nonumber
&P_{ij}(t,r_1,r_2;\gamma_1,\gamma_2)=P_j^\infty(r_1,r_2;\gamma_1,\gamma_2)\\\nonumber
&\hspace{3mm}+\sum_{l=2}^N\zeta_l(r_1,r_2;\gamma_1,\gamma_2)^t\times\\
&\hspace{3mm}\left\langle i\left|\psi_l(r_1,r_2;\gamma_1,\gamma_2)\right\rangle \left\langle\bar{\psi}_l(r_1,r_2;\gamma_1,\gamma_2)\right|j\right\rangle
\end{align}
can be inserted into the moments $\mathcal{R}_{ij}^{(0)}(r_{1},r_{2};\gamma_{1},\gamma_{2})$:
\begin{align}\nonumber
&\mathcal{R}_{ij}^{(0)}(r_{1},r_{2};\gamma_{1},\gamma_{2})\\\nonumber
&=\sum_{t=0}^\infty(P_{ij}(t,r_1,r_2;\gamma_1,\gamma_2)-P_j^\infty(r_1,r_2;\gamma_1,\gamma_2))\\
&=\sum_{l=2}^{N}\frac{\langle i|\psi_{l}(r_{1},r_{2};\gamma_{1},\gamma_{2})\rangle \langle \bar{\psi}_{l}(r_{1},r_{2};\gamma_{1},\gamma_{2})|j\rangle }{1-(1-\gamma_{2})(1-\gamma_{1})\lambda_{l}}.
 \label{eq:R_ij}
\end{align}
In this way, we can calculate the MFPT for the dynamics with two resetting nodes, by using the general Eq. (\ref{Tij_R}) valid for Markov processes
\begin{multline} \label{eq:MFPT_general}
    \langle T_{ij}(r_{1},r_{2};\gamma_{1},\gamma_{2})\rangle=\frac{\delta_{ij}}{P_{j}^{\infty}(r_{1},r_{2};\gamma_{1},\gamma_{2})}\\
+\frac{\mathcal{R}_{jj}^{(0)}(r_{1},r_{2};\gamma_{1},\gamma_{2})-\mathcal{R}_{ij}^{(0)}(r_{1},r_{2};\gamma_{1},\gamma_{2})}{P_{j}^{\infty}(r_{1},r_{2};\gamma_{1},\gamma_{2})}.
\end{multline}
\begin{figure*}[t!]
    \centering
    \includegraphics*[width=1.0\textwidth]{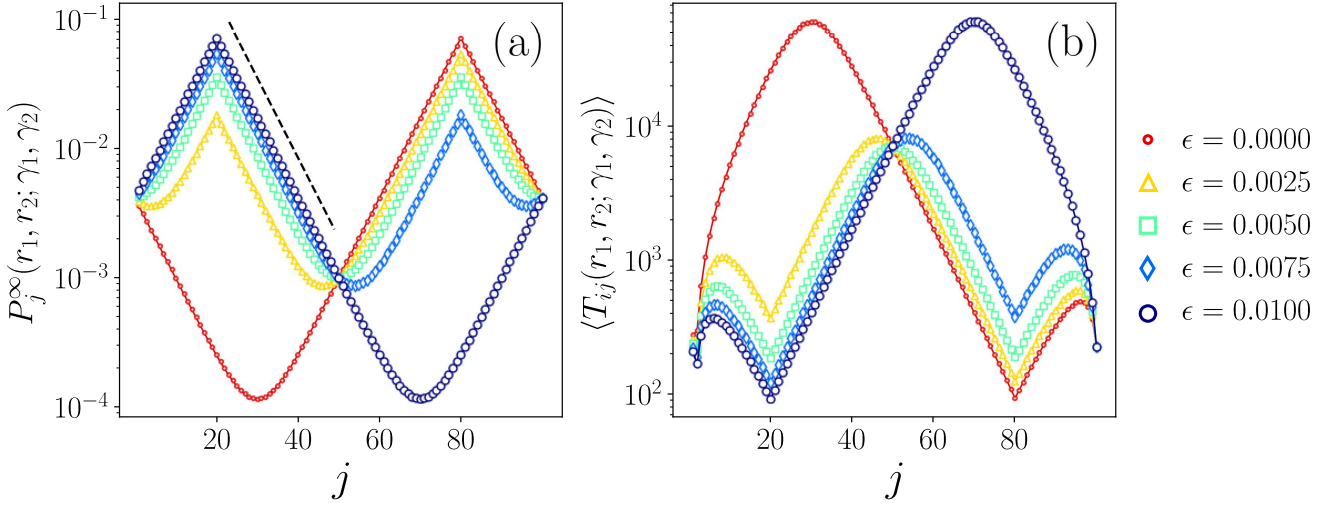}
    \vspace{-5mm}
    \caption{\label{Fig_2} Stationary distributions and mean first passage times for normal random walks with resetting to two nodes on a ring with $N = 100$. Stochastic restart to the nodes $r_{1}=20$ and $r_{2}=80$ occurs with probabilities $a_{1}=\epsilon$ and  $a_{2}=0.01-\epsilon$, respectively.  We show the effect of varying $\epsilon$ maintaining the total resetting probability $a_{1}+a_2=0.01$ constant.  (a) Stationary distribution $P_j^\infty(r_1,r_2;\gamma_1,\gamma_2) $ as a function of  $j$ obtained from Eq. \eqref{Pinf_resetM2_ring}, the dashed line represents the exponential decay $e^{-0.142\,d_{jr_1}}$. (b) MFPT $\langle T_{ij}(r_{1},r_{2};\gamma_{1},\gamma_{2})\rangle$ given by Eq. \eqref{MFPT_resetM2_ring} as a function of the target node $j$, for a walker initially at $i=1$.
}
\end{figure*}
The two factors that appear in the sum of Eq. \eqref{eq:R_ij} can be written as
\begin{align}\label{RijM2part_1}
    \langle i|\psi_{l}(r_{1},r_{2};\gamma_{1},\gamma_{2})\rangle&=\langle i | \phi_{l}\rangle -\kappa_{l}\langle r_{1}|\phi_{l}\rangle -\nu_{l}\langle r_{2}|\phi_{l}\rangle
\end{align}
and
\begin{equation}\label{RijM2part_2}
    \langle \bar{\psi}_{l}(r_{1},r_{2};\gamma_{1},\gamma_{2})|j\rangle = \langle \bar{\phi}_{l}|j\rangle.
\end{equation}
By substituting Eqs. (\ref{RijM2part_1})-(\ref{RijM2part_2}) into (\ref{eq:R_ij}), we obtain from Eq. \eqref{eq:MFPT_general} the mean first passage time 
\begin{multline} \label{eq:MFPT_vectors}
    \langle T_{ij}(r_{1},r_{2};\gamma_{1},\gamma_{2})\rangle =\frac{1}{P_{j}^{\infty}(r_{1},r_{2};\gamma_{1},\gamma_{2})}\times\\
    \bigg[\delta_{ij}+\sum_{l=2}^{N}\frac{\langle j|\phi_{l}\rangle \langle \bar{\phi}_{l}|j\rangle - \langle i | \phi_{l}\rangle \langle \bar{\phi}_{l}|j\rangle}{1-(1-\gamma_{2})(1-\gamma_{1})\lambda_{l}}\bigg].
\end{multline}
This relation is valid for any ergodic random walk and only depends on the eigenvalues and eigenvectors of the transition matrix without resetting, $\mathbf{W}$. In the following subsections, we apply these results to the analysis of local and non-local random walks on rings.
\subsection{Random walks with resetting on rings}
\label{Normal_Rings}
Let us consider an unbiased nearest-neighbor random walker with transition probabilities given by $w_{i\to j}=A_{ij}/k_i$, where $A_{ij}=A_{ji}=1$ if the two nodes are connected and 0 otherwise, and where $k_i=\sum_{l=1}^NA_{il}$ is the degree of $i$. With this transition matrix, transitions between nodes that are not directly connected to each other are not possible. We analyze the effects of resetting to two particular nodes on a ring with $N$ nodes. On this network, $k_i=2$ and the transition matrix $\mathbf{W}$ is a circulant matrix with well-known eigenvalues and eigenvectors \cite{VanMieghem2011,FractionalBook2019}.
\\[2mm]
These eigenvalues are given by $\lambda_{l}=\cos\varphi_{l}$, with $\varphi_{l}\equiv\frac{2\pi}{N}(l-1)$, whereas the projections of the eigenvectors in the canonical base are $\langle j|\phi_{l}\rangle=\frac{1}{\sqrt{N}}e^{-\mathrm{i}\varphi_{l}(j-1)}$ and $\langle \bar{\phi}_{l}|j\rangle=\frac{1}{\sqrt{N}}e^{\mathrm{i}\varphi_{l}(j-1)}$, where $\mathrm{i}=\sqrt{-1}$. For the problem of resetting to the nodes $r_1$ and $r_2$ with probabilities $a_1$ and $a_2$, respectively, we recall that $\gamma_1=a_1/(1-a_2)$ and $\gamma_2=a_2$. By using Eq. (\ref{eq:Pj_infty_final}), the exact expression for the stationary distribution reads
\begin{multline}
    P_j^\infty(r_1,r_2;\gamma_1,\gamma_2)=\frac{1}{N}\\+ \frac{1}{N}\sum_{m=2}^N\frac{\gamma_1 \cos(\varphi_{m}d_{jr_{1}})}{1-(1-\gamma_1)\cos\varphi_{m}}\times\\
    \bigg(1-\frac{\gamma_2}{1-(1-\gamma_2)(1-\gamma_1)\cos\varphi_{m}}\bigg)\\
+\frac{1}{N}\sum_{m=2}^N\frac{\gamma_2 \cos(\varphi_{m}d_{jr_{2}})}{1-(1-\gamma_2)(1-\gamma_1)\cos\varphi_m}, \label{Pinf_resetM2_ring}
\end{multline}
where $d_{ij}$ is the topological length of the shortest path connecting the nodes $i$ and $j$, therefore $\cos\left[\frac{2\pi}{N}d_{ij}\right]=\cos\left[\frac{2\pi}{N}(i-j)\right]$ for any pair $\{i,j\}$. Similarly, we obtain the MFPT from Eq. \eqref{eq:MFPT_vectors} with the corresponding eigenvalues and eigenvectors
\begin{multline}
    \langle T_{ij}(r_{1},r_{2};\gamma_{1},\gamma_{2})\rangle =\frac{1}{P_{j}^{\infty}(r_{1},r_{2};\gamma_{1},\gamma_{2})}\times\\
\bigg[\delta_{ij}
    + \sum_{l=2}^{N}\frac{1-\cos(\varphi_{l}d_{ij})}{1-(1-\gamma_{2})(1-\gamma_{1})\cos\varphi_{l}}\bigg].
    \label{MFPT_resetM2_ring}
\end{multline}
In Fig. \ref{Fig_2}, we present several examples obtained from the exact expressions (\ref{Pinf_resetM2_ring}) and (\ref{MFPT_resetM2_ring}) on a ring with $N=100$ nodes. Figure \ref{Fig_2}(a) displays the stationary distribution $P_{j}^{\infty}(r_{1},r_{2};\gamma_{1},\gamma_{2})$ as a function of the position $j$ for resetting to the nodes $r_1=20$ and $r_2=80$, with resetting probabilities $a_{1}=\epsilon$ and $a_{2}=0.01-\epsilon$, respectively. Hence the total resetting probability $a_1+a_{2}=0.01$ remains constant. The cases $\epsilon=0$ and $\epsilon=0.01$ reduce to a single-node resetting problem. Other values of $\epsilon$ produce two local maxima, at the resetting points $r_{1}$ and $r_{2}$, revealing a greater probability for the walker to be in these two points at large times.  
\\[2mm]
In Fig. \ref{Fig_2}(b) we depict the mean first passage times as a function of the position of the target node $j$, given the initial condition at $i=1$. It is interesting to notice that the shortest MFPT is attained when resetting occurs exclusively to the resetting node (say, $r_1$) that is closer to $j$. However, by introducing some amount of resetting to the other node (say, $r_2$), the MFPT to $j$ increases mildly, whereas the MFPTs at the nodes that are closer to $r_2$ decrease significantly, by more than one order of magnitude.  
\\[2mm]
To identify some basic properties of diffusive transport with resetting on rings, it is instructive to express Eqs. (\ref{Pinf_resetM2_ring}) and (\ref{MFPT_resetM2_ring}) in the limit of infinite rings. For readability, let us define  $b_{1}=1-\gamma_{1}$ and $b_{12}=(1-\gamma_{1})(1-\gamma_{2})$. Therefore, by taking the limit $N\to\infty$ and  considering $d\varphi=\frac{2\pi}{N}$, the stationary distribution becomes
\begin{multline}\label{PinfOnfRing_M2_integral}
 P_j^\infty(r_1,r_2;\gamma_1,\gamma_2)= \frac{\gamma_1}{2\pi}\int_0^{2\pi}\frac{\cos(\varphi d_{jr_{1}})}{1-b_1\cos\varphi}d\varphi\\
+\frac{\gamma_2}{2\pi}\int_0^{2\pi}\frac{\cos(\varphi d_{jr_{2}})}{1-b_{12}\cos\varphi}d\varphi\\
-\frac{\gamma_1\gamma_2}{2\pi}\int_0^{2\pi}\frac{\cos(\varphi d_{jr_{1}})}{(1-b_1\cos\varphi)(1-b_{12}\cos\varphi)}d\varphi.
\end{multline}
By using the identity \cite{ResetNetworks_PRE2020}
\begin{equation}
\frac{1}{2\pi}\int_0^{2\pi} \frac{\cos(x\theta)}{1-b\cos(\theta)}d\theta=\frac{\left(\frac{1+\sqrt{1-b^2}}{b}\right)^{-x}}{\sqrt{1-b^2}}
\end{equation}
we deduce (see Appendix \ref{Appendix_PartA} for details)
\begin{align}\nonumber
&\frac{1}{2\pi}\int_{0}^{2\pi}\frac{\cos(\varphi x)}{(1-y\cos(\varphi))(1-z\cos(\varphi))}d\varphi\\
&=\frac{y}{y-z}\frac{\left(\frac{1+\sqrt{1-y^2}}{y}\right)^{-x}}{\sqrt{1-y^2}}-
\frac{z}{y-z}\frac{\left(\frac{1+\sqrt{1-z^2}}{z}\right)^{-x}}{\sqrt{1-z^2}},
\label{Integral_InfLine_M2}
\end{align}
and, after simplifications, 
\begin{multline}
    P_{j}^{\infty}(r_{1},r_{2};\gamma_{1},\gamma_{2})=
    \frac{\gamma_{1}(1-\gamma_{2})}{\sqrt{1-b_{12}^2}}\Bigg(\frac{1+\sqrt{1-b_{12}^2}}{b_{12}}\Bigg)^{-d_{jr_{1}}}\\
    +\frac{\gamma_{2}}{\sqrt{1-b_{12}^2}}\bigg(\frac{1+\sqrt{1-b_{12}^2}}{b_{12}}\bigg)^{-d_{jr_{2}}}.
\end{multline}
By defining $\chi\equiv\ln\left[\frac{1+\sqrt{1-b_{12}^2}}{b_{12}}\right]$, or equivalently, 
\begin{equation}
\chi\equiv\ln\left[\frac{1+\sqrt{1-(1-a_1-a_2)^2}}{1-a_1-a_2}\right],
\end{equation}
we obtain the rather simple expression
\begin{equation}
    \label{eq:stat_dist_lim_final}
    P_{j}^{\infty}(r_{1},r_{2};\gamma_{1},\gamma_{2})=
    \frac{\gamma_{1}(1-\gamma_{2})e^{-\chi d_{jr_{1}}}+\gamma_{2}e^{-\chi d_{jr_{2}}}}{\sqrt{1-(1-\gamma_{1})^2(1-\gamma_{2})^2}}.
\end{equation}
If either $\gamma_{1}$ or $\gamma_{2}$ are set to zero in Eq. (\ref{eq:stat_dist_lim_final}), we recover the non-equilibrium steady state for resetting to a single node derived in Ref. \cite{ResetNetworks_PRE2020}.
\\[2mm]
By applying the same procedure to Eq. (\ref{eq:MFPT_vectors}), the MFPT for an infinite ring takes the form
\begin{multline}
    \langle T_{ij}(r_{1},r_{2};\gamma_{1},\gamma_{2})\rangle =\frac{\delta_{ij}}{P_{j}^{\infty}(r_{1},r_{2};\gamma_{1},\gamma_{2})}\\
    +\frac{1}{P_{j}^{\infty}(r_{1},r_{2};\gamma_{1},\gamma_{2})} \int_{0}^{2\pi}\frac{1-\cos(\varphi d_{ij})}{1-b_{12}\cos(\varphi)}d\varphi.
\end{multline}
In particular, the mean first return time to the starting site (setting $j=i$) reads
\begin{equation}
    \langle T_{ii}(r_{1},r_{2};\gamma_{1},\gamma_{2})\rangle =\frac{1}{P_{j}^{\infty}(r_{1},r_{2};\gamma_{1},\gamma_{2})},
\end{equation}
in agreement with Kac's lemma \cite{kac1947notion}. For $i\neq j$, we get
\begin{multline}
    \langle T_{ij}(r_{1},r_{2};\gamma_{1},\gamma_{2})\rangle = \frac{1}{P_{j}^{\infty}(r_{1},r_{2};\gamma_{1},\gamma_{2})}\times\\
\frac{1}{\sqrt{1-b_{12}^{2}}}\bigg[1-\bigg(\frac{1+\sqrt{1-b_{12}^2}}{b_{12}}\bigg)^{-d_{ij}}\bigg].
\end{multline}
or,
\begin{equation}\label{TijInfRing_M2}
 \langle T_{ij}\rangle= \frac{1-e^{-\chi d_{ij}}}{\gamma_{1}(1-\gamma_{2})e^{-\chi d_{jr_{1}}}+\gamma_{2}e^{-\chi d_{jr_{2}}}}.
\end{equation}
Clearly, the behaviors of the stationary state and of the MFTP $\langle T_{ij}(r_{1},r_{2};\gamma_{1},\gamma_{2})\rangle$ with respect to $d_{jr_1}$, $d_{jr_2}$ or $d_{ij}$, are controlled by the characteristic length-scale $\chi^{-1}$, which depends only on the non-resetting probability $b_{12}=1-a_1-a_2$.
\\[2mm]
The exact results in Eqs. (\ref{eq:stat_dist_lim_final})  and (\ref{TijInfRing_M2}) for the infinite ring help us to understand the exponential behavior observed in  Fig. \ref{Fig_2}(a) around the nodes $r_1$ and $r_2$. Since $a_1+a_2$ is fixed to $0.01$ in all these examples, we have $b_{12}=0.99$ or $\chi\approx 0.142$. Therefore, close to the nodes $r_1$ and $r_2$ the stationary distributions decay exponentially with respect to the distances $d_{jr_1}$ or $d_{jr_2}$, with the same slope $\chi$ represented by a dashed line in  Fig. \ref{Fig_2}(a).
\subsection{L\'evy flights with resetting on rings}
\label{Levy_Rings}
Let us now explore the effects of resetting on a random walker with long-range steps. 
L\'evy flights on an arbitrary graph can be generated by taking a power of the Laplacian matrix $\mathbf{L}$, with elements $L_{ij}=\delta_{ij} k_i-A_{ij}$. Let us introduce the transition probabilities \cite{RiascosMateosFD2014}
\begin{equation}\label{wijfrac}
w_{i\to j}(\alpha)=\delta_{ij}-\frac{(\mathbf{L}^\alpha)_{ij}}{(\mathbf{L}^\alpha)_{ii}}\qquad 0<\alpha< 1.
\end{equation}
In particular, for $\alpha\to 1$ one recovers the simple random walk with transitions to nearest-neighbor nodes. The transition probabilities in Eq. (\ref{wijfrac}) with $\alpha<1$ define a L\'evy flight with $w_{i\to j}(\alpha)\sim d_{ij}^{-(1+2\alpha)}$, where the distance $d_{ij}$ is the length of the shortest path on the graph between $i$ and $j$, and where $d_{ij}\gg1$ (see Refs. \cite{RiascosMateos2012,RiascosMateosFD2015,FractionalBook2019} for a detailed discussion on L\'evy flights and fractional transport on networks).
\\[2mm]
In the case of finite rings with $N$ nodes, the Laplacian $\mathbf{L}$ as well as the fractional Laplacian $\mathbf{L}^\alpha$ and the matrix with the elements given by Eq. (\ref{wijfrac}) are circulant matrices \cite{RiascosMateosFD2015,FractionalBook2019}. Therefore the left and right eigenvectors are the same as the ones exposed in Section \ref{Normal_Rings}. However, the eigenvalues $\{\lambda_l(\alpha)\}_{l=1} ^N$ of the transition matrix defined in Eq. (\ref{wijfrac}) are given by \cite{RiascosMateosFD2015,RiascosMichelitsch2017_gL}
\begin{equation}\label{lambdaGLring}
\lambda_l(\alpha)=1-\frac{1}{k^{(\alpha)}}\, \left(2-2\cos\varphi_l\right)^\alpha
\end{equation}
where $\varphi_l=\frac{2\pi}{N}(l-1)$ and the {\it fractional degree} $k^{(\alpha)}$ is defined as \cite{RiascosMateosFD2015,RiascosMichelitsch2017_gL}
\begin{equation}\label{degreeGLring}
k^{(\alpha)}=\frac{1}{N}\sum_{l=1}^N \left(2-2\cos\varphi_l\right)^\alpha.
\end{equation}
In the case of L\'evy flights on rings with resetting to a single node, we can use Eq. (\ref{Pinfvectors_1}) with the eigenvalues (\ref{lambdaGLring}) and obtain the stationary distribution
\begin{align}\nonumber
P_j^\infty(i;\gamma)&=\frac{1}{N}+\gamma\sum_{l=2}^N\frac{\left\langle i|\phi_l\right\rangle \left\langle\bar{\phi}_l|j\right\rangle}{1-(1-\gamma)\lambda_l(\alpha)}\\\nonumber
&=\frac{1}{N}+\frac{\gamma}{N}\sum_{l=2}^N\frac{e^{-\mathrm{i}\frac{2\pi(l-1)(i-j)}{N}}}{1-(1-\gamma)\lambda_l(\alpha)}\\
&=\frac{1}{N}+\frac{\gamma}{N}\sum_{l=2}^N \frac{\cos\left(\varphi_l\,d_{ij}\right)}{1-(1-\gamma)\lambda_l(\alpha)},
\label{Pinf_levyfinite_ring}
\end{align}
which is valid for $0<\alpha\leq 1$. Similarly, from Eq. (\ref{MFPT_resetSM}) with $r=i$ we obtain the MFPT for L\'evy flights with reset to the initial node
\begin{align}\nonumber
&\left\langle T_{ij}(i;\gamma)\right\rangle=\frac{\delta_{ij}+
	\sum\limits_{l=2}^N\frac{\left\langle j|\phi_l\right\rangle \left\langle\bar{\phi}_l|j\right\rangle-\left\langle i|\phi_l\right\rangle \left\langle\bar{\phi}_l|j\right\rangle
	}{1-(1-\gamma)\lambda_l(\alpha)}}{P_j^\infty(i;\gamma)}\\\nonumber
&\qquad\qquad=\frac{\delta_{ij}+\frac{1}{N}
	\sum\limits_{l=2}^N\frac{1-e^{-\mathrm{i}\frac{2\pi(l-1)(i-j)}{N}}}{1-(1-\gamma)\lambda_l(\alpha)}}{P_j^\infty(i;\gamma)}\\
&=\frac{1}{P_j^\infty(i;\gamma)}\left[\delta_{ij}+\frac{1}{N}
\sum_{l=2}^N\frac{1-\cos\left(\varphi_l\,d_{ij}\right)}{1-(1-\gamma)\lambda_l(\alpha)}\right]. \label{MFPT_levy_ring_finite}
\end{align}
The expressions corresponding to L\'evy flights with resetting to two nodes are deduced in a similar way, by replacing the terms $\cos(\varphi_l)$ by the generalized eigenvalues $\lambda_l(\alpha)$ in the relations (\ref{Pinf_resetM2_ring}) and (\ref{MFPT_resetM2_ring}).
\begin{figure*}[t!]
    \centering
    \includegraphics*[width=0.95\textwidth]{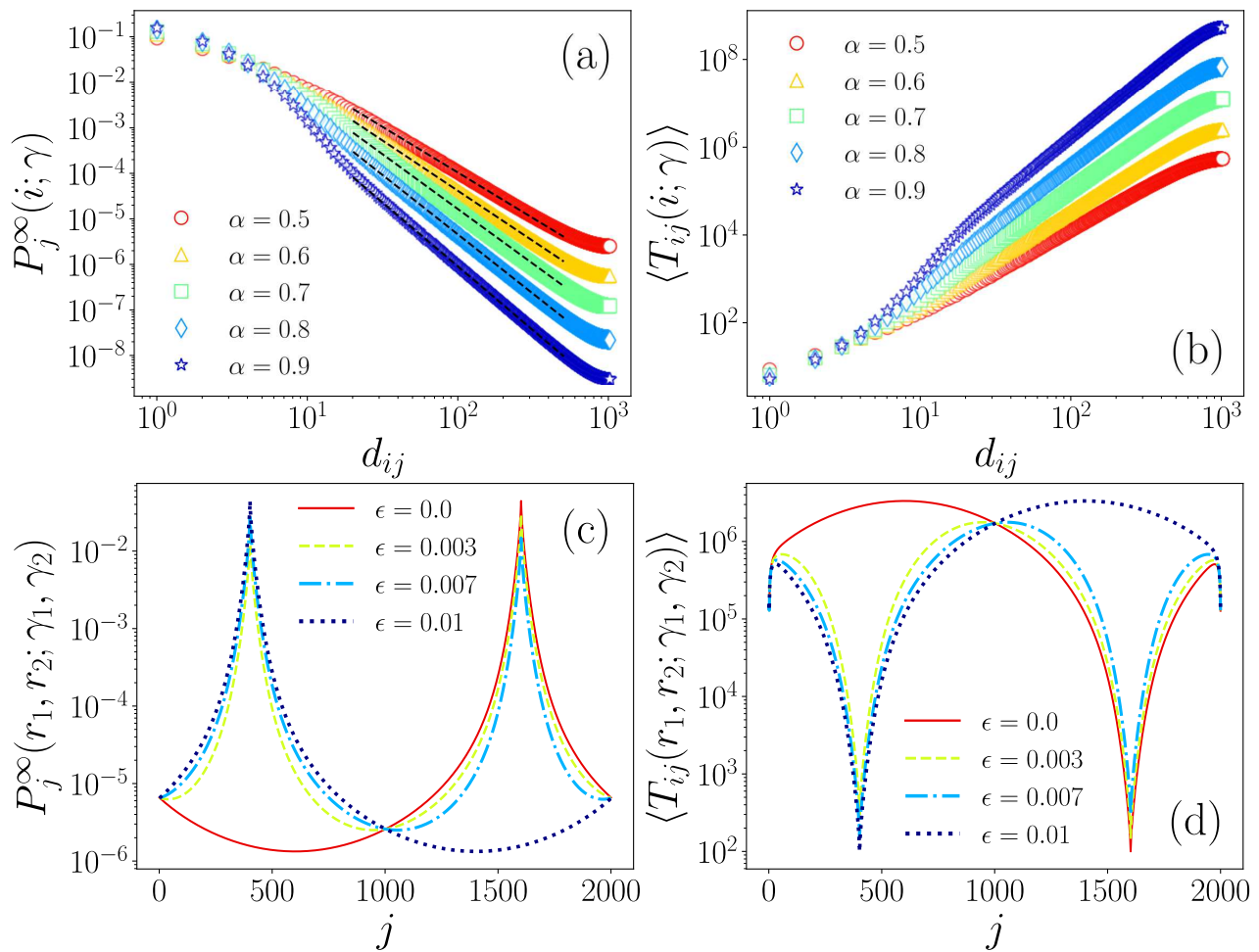}
    \vspace{-3mm}
    \caption{\label{Fig_3} Stationary distributions and MFPTs for L\'evy flights with resetting to one and two nodes, on a ring with $N=2000$. For resetting to one node, we show (a) $P_j^\infty(i;\gamma)$ and (b) $\left\langle T_{ij}(i;\gamma)\right\rangle$ as a function of the distance $d_{ij}$ between the initial node $i$, where resetting occurs, and the node $j$ for different values of $\alpha$. The resetting probability is $\gamma=0.2$. The dashed lines in (a) represent power-laws $\propto d_{ij}^{-(1+2\alpha)}$. For resetting to two nodes: (c) $P_j^\infty(r_1,r_2;\gamma_1,\gamma_2) $ and (d) MFPT $\langle T_{ij}(r_{1},r_{2};\gamma_{1},\gamma_{2})\rangle$ as a function of $j$, considering the initial node $i=1$. In these cases $\alpha=0.75$ is constant and the resetting probabilities to the nodes $r_1=400$ and $r_2=1600$ are $a_{1}=\epsilon$ and $a_{2}=0.01-\epsilon$, respectively.
}
\end{figure*}
\\[2mm]
In Figs. \ref{Fig_3}(a)-(b) we analyze the dynamics with resetting probability $\gamma$ solely to the initial node ($i=r$), on a ring with $N=2000$.  Fig. \ref{Fig_3}(a) displays the numerical values of $P_j^\infty(i;\gamma)$ obtained from  Eq. (\ref{Pinf_levyfinite_ring}) as a function of the distance $d_{ij}$ for $\alpha=0.5,0.6,\ldots,0.9$ . The stationary steady state exhibits a power-law behavior $P_j^\infty(i;\gamma)\propto d_{ij}^{-(1+2\alpha)}$ for $1\ll d_{ij}\ll N/2$ and $1/2<\alpha< 1$ [see the dashed lines in  Fig. \ref{Fig_3}(a)]. This asymptotic behavior is explored analytically for L\'evy flights on infinite rings in Appendix \ref{Appendix_PartB}, and is consistent with the findings of Ref. \cite{kusmierz2015optimal} on continuous L\'evy flights under resetting on the infinite line.  The scaling-law contrasts with the exponential decay of the occupation probability of a simple random walk, $P_j^\infty (i;\gamma)\approx\frac{\sqrt{2\gamma }}{2}e^{-\sqrt{2\gamma }d_{ij}}$ when the resetting probability $\gamma$ is small \cite{ResetNetworks_PRE2020}. In Fig. \ref{Fig_3}(b) we present the MFPT given by Eq. (\ref{MFPT_levy_ring_finite}).
\\[2mm]
Likewise, Figs. \ref{Fig_3}(c)-(d) show the results for resetting to two nodes. The results show the same qualitative features as in our previous analysis of the local dynamics in Fig. \ref{Fig_2}. We considered L\'evy flights with $\alpha=0.75$ and resetting to the nodes $r_1=400$ and $r_2=1600$, with probabilities $a_1=\epsilon$ and $a_2=0.01-\epsilon$, thus maintaining the total resetting probability $a_1+a_2$ constant.
\section{Random walks with resetting to multiple nodes}
\label{RW_resetM}
In this section, we extend the methods introduced above to consider the resetting to $\mathcal{M}$ different nodes. We apply this approach to analyze simple random walks on a comb graph, L\'evy flights that visit points on a two-dimensional region and Google's random walks on regular networks.
\subsection{General results}
\label{RW_resetM_general}
We consider $\mathcal{M}$ nodes $r_1,r_2,\ldots,r_\mathcal{M}$ to which resetting is performed stochastically with probabilities
$0 \leq a_s\leq 1$ (where $s=1,\ldots, \mathcal{M}$). The transition matrix of the random walk with resetting to these nodes reads
\begin{multline}\label{matPi_Mreset}
\mathbf{\Pi}(r_1,r_2,\ldots,r_\mathcal{M};a_1,a_2,\ldots,a_\mathcal{M})\\
=a_0\mathbf{W}+\sum_{s=1}^\mathcal{M} a_s\mathbf{\Theta}(r_s),
\end{multline}
with the same notation as before, where $ a_0\equiv 1-\sum_{s=1}^\mathcal{M}a_s$. We assume that the total resetting probability $\beta\equiv\sum_{s=1}^\mathcal{M}a_s$ is such that $0\leq \beta < 1$.  
\\[2mm]
Let us introduce the simplified notation for the transition matrix in Eq. (\ref{matPi_Mreset}),
\begin{equation}
\mathbf{\Pi}(r_1,r_2,\ldots,r_\mathcal{M};a_1,a_2,\ldots,a_\mathcal{M})=\mathbf{\Pi}_\mathcal{M}.
\end{equation}
The matrix $\mathbf{\Pi}_\mathcal{M}$ can be  obtained iteratively through the relation
\begin{equation}
\mathbf{\Pi}_s=\frac{\sum_{l=0}^{s-1}a_l}{\sum_{l=0}^{s}a_l}\mathbf{\Pi}_{s-1}+\frac{a_s}{\sum_{l=0}^{s}a_l}\mathbf{\Theta}(r_s),
\end{equation}
for $s=1,2,\ldots, \mathcal{M}$ with $\mathbf{\Pi}_0\equiv\mathbf{W}$. We also define
\begin{equation}\label{gamma_gendef}
\gamma_s\equiv \frac{a_s}{\sum_{l=0}^{s}a_l},\qquad s=1,2,\ldots,\mathcal{M}.
\end{equation}
Therefore $\frac{\sum_{l=0}^{s-1}a_l}{\sum_{l=0}^{s}a_l}=1-\gamma_s$ and
\begin{equation}
\mathbf{\Pi}_s=(1-\gamma_s)\mathbf{\Pi}_{s-1}+\gamma_s\mathbf{\Theta}(r_s).
\end{equation}
In the following, we use the compact notation  $|\psi^{(s)}_{l}\rangle$  and  $\langle \bar{\psi}^{(s)}_{l}|$  for the right and left eigenvectors of $\mathbf{\Pi}_s$ corresponding to the eigenvalue $\zeta_l^{(s)}$. According to Eq. (\ref{eigvals_zeta}),
the eigenvalues satisfy $\zeta_1^{(s)}=1$ and
\begin{align}\nonumber
\zeta_l^{(s)}&=(1-\gamma_s)\zeta_l^{(s-1)}=(1-\gamma_s)(1-\gamma_{s-1})\zeta_l^{(s-2)}\\
&=\cdots=\lambda_l\prod_{m=1}^s(1-\gamma_m),\qquad l=2,3,\ldots, N.\label{eigval_reset_ws}
\end{align}
By construction, $\zeta_l^{(0)}=\lambda_l$, which are the eigenvalues of $\mathbf{W}$. For the eigenvectors, we have
\begin{equation}
\label{rpsi_reset_1}
|\psi^{(s)}_{1}\rangle = |\psi^{(s-1)}_{1}\rangle= |\psi^{(s-2)}_{1}\rangle=\cdots=|\psi^{(0)}_{1}\rangle=|\phi_{1}\rangle,
\end{equation}
and, for $m=2,\ldots,N$
\begin{equation}
\label{lpsi_reset_s}
\langle\bar{\psi}^{(s)}_m|=\langle\bar{\psi}^{(s-1)}_m|=\cdots
=\langle\bar{\psi}^{(0)}_m|=\langle\bar{\phi}_m|.
\end{equation}
These two classes of eigenstates remain unaltered with the introduction of resetting. Conversely, combining Eqs. (\ref{psirl_reset}) for $l=2,\ldots,N$ and (\ref{rpsi_reset_1}), yields
\begin{equation}\label{psi_s_Mnodes}
    |\psi^{(s)}_{l}\rangle = |\psi^{(s-1)}_{l}\rangle-\frac{\gamma_s}{1-\zeta_l^{(s)}}\frac{\langle r_{s}|\psi^{(s-1)}_{l}\rangle}{\langle r_{s}|\phi_{1}\rangle}|\phi_{1}\rangle.
\end{equation}
For the remaining $\langle\bar{\psi}^{(s)}_1|$, one combines Eq. (\ref{psil1}) with (\ref{rpsi_reset_1}) and (\ref{lpsi_reset_s}) to obtain
\begin{equation}\label{bar_psi_s}
\langle\bar{\psi}^{(s)}_1|=\langle\bar{\psi}^{(s-1)}_1|
+\sum_{m=2}^N\frac{\gamma_s}{1-\zeta_l^{(s)}}\frac{\langle r_s|\psi^{(s-1)}_m\rangle}{\left\langle r_s|\phi_1\right\rangle}\left\langle\bar{\phi}_m\right|.
\end{equation}
In terms of these eigenvectors, the stationary distribution is given by
\begin{align}\nonumber
P_{j}^{\infty}(\vec{r};\vec{\gamma})&\equiv P_{j}^{\infty}(r_{1},r_{2},\ldots,r_{\mathcal{M}};\gamma_{1},\gamma_{2},\ldots,\gamma_{\mathcal{M}})\\
&=\langle j|\psi^{(\mathcal{M})}_{1}\rangle \langle\bar{\psi}^{(\mathcal{M})}_1| j\rangle=\langle j|\phi_{1}\rangle \langle\bar{\psi}^{(\mathcal{M})}_1| j\rangle.
\end{align}
The stationary distribution $P_{j}^{\infty}(\vec{r};\vec{\gamma})$ is thus obtained from the eigenvalues and eigenvectors of the transition matrix without resetting $\mathbf{W}$ through iterations of Eqs. (\ref{eigval_reset_ws})-(\ref{bar_psi_s}) until $\langle\bar{\psi}^{(\mathcal{M})}_1| j\rangle$ is retrieved. As this approach allows us to calculate all the eigenvalues and eigenvectors of $\mathbf{\Pi}_{\mathcal{M}}$, these can be used to compute the MFPT of the resetting process to the $\mathcal{M}$ nodes. The moments of the occupation probability $P_{ij}(t;\vec{r},\vec{\gamma})$ that appear in the general relation (\ref{Tij_R}) are given by
\begin{align}\nonumber
\mathcal{R}^{(0)}(i,j;\vec{r},\vec{\gamma})&=\sum_{t=0}^\infty \left[ P_{ij}(t;\vec{r},\vec{\gamma})-P_j^\infty(\vec{r};\vec{\gamma}) \right]\\ \nonumber
&=\sum_{t=0}^\infty \sum_{l=2}^N (\zeta_{l}^{(\mathcal{M})})^t \langle i |\psi^{(\mathcal{M})}_l\rangle\langle\bar{\psi}^{(\mathcal{M})}_l|j\rangle\\
&=\sum_{l=2}^N  \frac{1}{1-\zeta_{l}^{(\mathcal{M})}} \langle i |\psi^{(\mathcal{M})}_l\rangle\langle\bar{\psi}^{(\mathcal{M})}_l|j\rangle.\label{Rij_Mnodes_Psi}
\end{align}
Hence,
\begin{align}\nonumber
&\mathcal{R}^{(0)}(j,j;\vec{r},\vec{\gamma})-\mathcal{R}^{(0)}(i,j;\vec{r},\vec{\gamma})\\
&=\sum_{l=2}^N  \frac{\langle j |\psi^{(\mathcal{M})}_l\rangle\langle\bar{\psi}^{(\mathcal{M})}_l|j\rangle-\langle i |\psi^{(\mathcal{M})}_l\rangle\langle\bar{\psi}^{(\mathcal{M})}_l|j\rangle}{1-\zeta_{l}^{(\mathcal{M})}}.
\end{align}
However, from Eqs. (\ref{lpsi_reset_s})-(\ref{psi_s_Mnodes}), we obtain
\begin{align}\nonumber
&\langle j|\psi^{(\mathcal{M})}_{l}\rangle \langle\bar{\psi}^{(\mathcal{M})}_l| j\rangle-\langle i|\psi^{(\mathcal{M})}_{l}\rangle \langle\bar{\psi}^{(\mathcal{M})}_l| j\rangle\\\nonumber
&=\langle j|\psi^{(\mathcal{M}-1)}_{l}\rangle \langle\bar{\psi}^{(\mathcal{M}-1)}_l| j\rangle-\langle i|\psi^{(\mathcal{M}-1)}_{l}\rangle \langle\bar{\psi}^{(\mathcal{M}-1)}_l| j\rangle\\ \nonumber
&=\langle j|\psi^{(\mathcal{M}-2)}_{l}\rangle \langle\bar{\psi}^{(\mathcal{M}-2)}_l| j\rangle-\langle i|\psi^{(\mathcal{M}-2)}_{l}\rangle \langle\bar{\psi}^{(\mathcal{M}-2)}_l| j\rangle\\\nonumber
&=\cdots\\\nonumber
&=\langle j|\psi^{(0)}_{l}\rangle \langle\bar{\psi}^{(0)}_l| j\rangle-\langle i|\psi^{(0)}_{l}\rangle \langle\bar{\psi}^{(0)}_l| j\rangle\\\nonumber
&=\langle j|\phi_{l}\rangle \langle\bar{\phi}_l| j\rangle-\langle i|\phi_{l}\rangle \langle\bar{\phi}_l| j\rangle.
\end{align}
Therefore
\begin{figure*}[!t]
\begin{center}
\includegraphics*[width=0.95\textwidth]{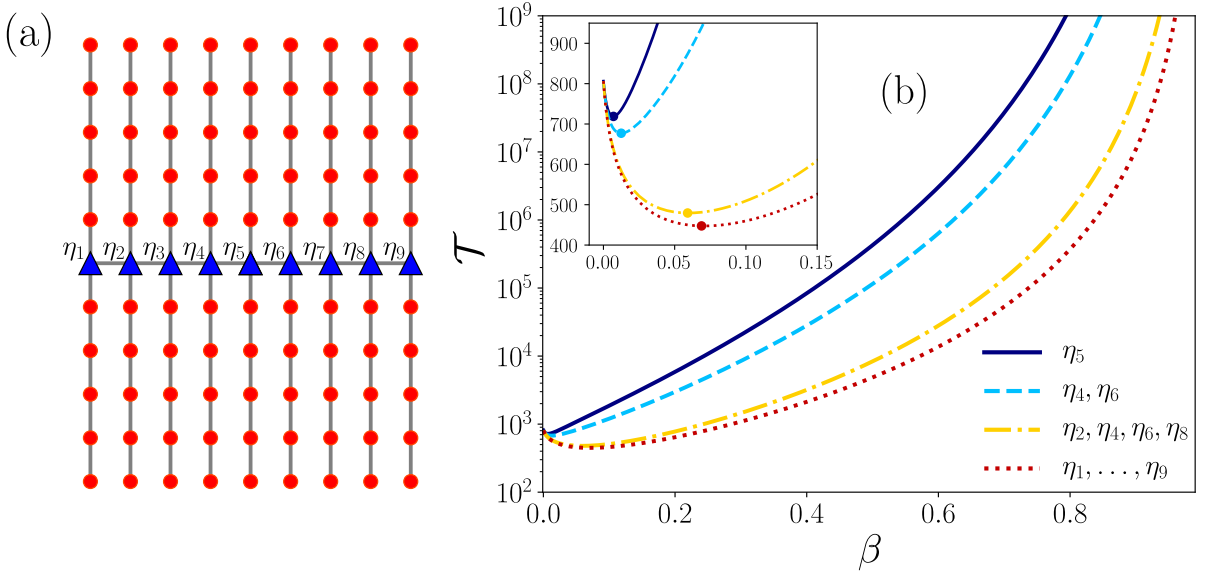}
\end{center}
\vspace{-4mm}
\caption{\label{Fig_4} Random walks with multiple resetting on a comb graph with $N=99$ nodes. (a) Comb graph: the nodes in the central line are represented with triangles and denoted as $\eta_1, \eta_2,\ldots,\eta_9$. (b) Global time $\mathcal{T}$ as a function of the total resetting probability $\beta$; four cases are explored: resetting with probability $\beta$ to the central node $\eta_5$ only, resetting with probability $\beta/2$ to the nodes $\eta_4,\eta_6$, resetting with probability $\beta/4$ to the nodes $\eta_2,\eta_4,\eta_6,\eta_8$, and resetting with probability $\beta/9$ to all the nodes of the central line. The inset shows the results for $0\leq\beta\leq 0.15$ and the minimum global time $\mathcal{T}^\star$ for each curve is represented with a circle.}
\end{figure*}
\begin{multline}\label{RjjRij_reset}
\mathcal{R}^{(0)}(j,j;\vec{r},\vec{\gamma})-\mathcal{R}^{(0)}(i,j;\vec{r},\vec{\gamma})
\\=
\sum_{l=2}^N  \frac{1}{1-\zeta_{l}^{(\mathcal{M})}} \left[\langle j|\phi_{l}\rangle \langle\bar{\phi}_l| j\rangle-\langle i|\phi_{l}\rangle \langle\bar{\phi}_l| j\rangle\right].
\end{multline}
This relation shows that the net effect of having multiple resetting points in the expression $\mathcal{R}^{(0)}(j,j;\vec{r},\vec{\gamma})-\mathcal{R}^{(0)}(i,j;\vec{r},\vec{\gamma})$ lies in the modification of the eigenvalues. Therefore, applying Eq. (\ref{Tij_R}), which is valid for ergodic random walks and an arbitrary number of resetting nodes, leads to the MFPT
\begin{multline}\label{MFPT_Mreset}
    \langle T_{ij}(\vec{r};\vec{\gamma})\rangle =\frac{\delta_{ij}}{P_{j}^{\infty}(\vec{r};\vec{\gamma})}\\+\frac{1}{P_{j}^{\infty}(\vec{r};\vec{\gamma})} \sum_{l=2}^{N}\frac{\langle j|\phi_{l}\rangle \langle \bar{\phi}_{l}|j\rangle - \langle i | \phi_{l}\rangle \langle \bar{\phi}_{l}|j\rangle}{1-z(\vec{\gamma})\lambda_{l}}
\end{multline}
with $z(\vec{\gamma})\equiv \prod_{s=1}^\mathcal{M}(1-\gamma_s)$.
\\[2mm]
Having at hand the analytical expressions for the MFPT of a random walk with resetting to $\mathcal{M}$ nodes, it is also convenient to define a \emph{global} first passage time obtained from averaging that quantity over all the starting and target nodes
\begin{equation}\label{Global_Tdef}
\mathcal{T}=\frac{1}{N^2}\sum_{i=1}^N\sum_{j=1}^N\langle T_{ij}\rangle.
\end{equation}
This global time quantifies the capacity of a random walk strategy to quickly reach any node of the network from any other node.
\subsection{Random walks on a comb graph}
In Fig. \ref{Fig_4}, we explore the effects of different types of resetting on normal random walks on a comb graph with $N=99$. Comb graphs are branched structures obtained from a linear graph with $L_x$ nodes by attaching to each node two side chains of length $L_y/2$ \cite{AgliariPRE2016,AgliariPRE2019}. The resulting structure is a tree with $N=L_x( L_y+1)$ nodes. In Fig. \ref{Fig_4}(a) we present a network with $L_x=9$ and $L_y=10$, the $L_x$ nodes in the central linear graph are represented with triangles and denoted as $\eta_1,\ldots,\eta_9$.
\\[2mm]
In Fig. \ref{Fig_4}(b), we show the global time $\mathcal{T}$ corresponding to this network as a function of the total resetting probability $\beta$. We apply the general approach of Section \ref{RW_resetM_general} to calculate the mean first passage times (\ref{MFPT_Mreset}) in the case of a nearest-neighbor random walk with resetting to different nodes in the central line of the comb graph.  We analyze four cases: resetting to the central node $\eta_5$ only, with probability $\beta$; resetting to the nodes $\eta_4$ and $\eta_6$ with probabilities $\beta/2$ each; resetting to the four nodes  $\eta_2,\eta_4,\eta_6,\eta_8$ with probabilities $\beta/4$ each; and resetting to all the nodes of the central line $\eta_1,\ldots,\eta_9$, each with probability $\beta/9$. Our findings show that adding more resetting nodes from the central line accelerates the exploration of the network. The behavior of $\mathcal{T}$ with $\beta$ is non-monotonic in each case, and the value $\beta^\star$ that minimizes the global time increases with the number of resetting nodes. The minimal values of $\mathcal{T}$ are obtained when we consider the resetting to all the nodes in the central line.
\subsection{Long-range dynamics on continuous spaces}
In this part, we explore the ability of a random walk under single or multiple resetting to reach specific locations in space. This problem is inspired by the modeling of human mobility in urban areas with agents visiting points of interest  \cite{RiascosMateosPlos2017,LoaizaMonsalvePlosOne2019,RiascosMateosSciRep2020}. The type of motion considered here is an example of random walks taking place in a continuous space but modeled with the formalism of random walks on networks.
\\[2mm]
Let us consider $N$ points (locations) in a $2D$ plane labeled as $i=1,2, \ldots ,N$. The coordinates of each point are arbitrary and $l_{ij}$ is the Euclidean distance between $i$ and $j$ (the distance $l_{ij}=l_{ji}\geq 0$ can also be calculated using other metrics \cite{RiascosMateosPlos2017}). We define a discrete time random walker (without resetting) that visits at each step one of these locations according to the transition probability $w_{i\to j}^{(\alpha)}(R)$ \cite{RiascosMateosPlos2017}
\begin{figure*}[!t]
\begin{center}
\includegraphics*[width=0.95\textwidth]{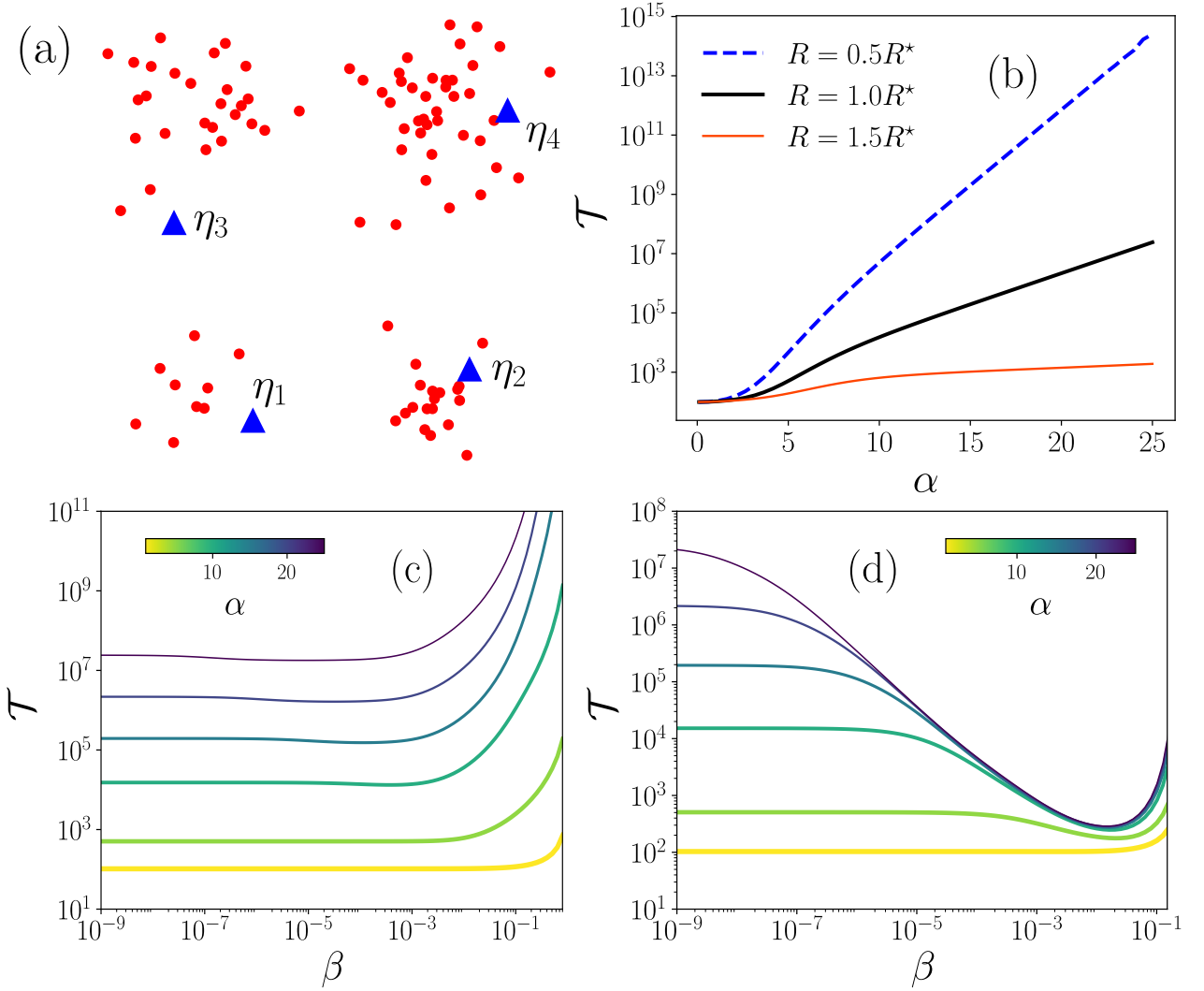}
\end{center}
\vspace{-5mm}
\caption{\label{Fig_5} Random walks visiting $N=100$ points. (a) Distribution of points in the domain $[0,1]\times [0,1]$, the points $\eta_m$ with $m=1,2,3,4$ represent possible resetting locations. (b) Global time $\mathcal{T}$ for the L\'evy flight model without resetting given by the transition matrix (\ref{wijRalpha})-(\ref{Omega_ij}) and for different values of $\alpha$ and $R$, $R^\star\equiv \sqrt{\frac{\ln N}{\pi N}}$ is a reference length. (c)  $\mathcal{T}$ for the same model under resetting to the single node $\eta_4$, located in the largest cluster, as a function of the resetting probability $\beta$. (d)  $\mathcal{T}$ as a function of the total resetting probability $\beta$ for the model under resetting to the four points $\eta_1,\ldots,\eta_4$, with probability $\beta/4$ each. In (c) and (d), we take $R=R^\star$ and $\alpha=1,5,10,15,20,25$ (codified in the colorbars).
}
\end{figure*}
\begin{equation}\label{wijRalpha}
w_{i\to j}^{(\alpha)}(R)=\frac{\Omega_{ij}^{(\alpha)}(R)}{\sum_{m=1}^{N} \Omega_{im}^{(\alpha)}(R)},
\end{equation}
where the weights $\Omega_{ij}^{(\alpha)}(R)$ are defined by \cite{RiascosMateosPlos2017}
\begin{eqnarray}\label{Omega_ij}
\Omega_{ij}^{(\alpha)}(R)&=
\left\{
\begin{array}{ll}
0 & \mathrm{if} \qquad i=j,\\
1 & \mathrm{for} \quad  0< l_{ij}\leq R,\\
\left(R/l_{ij}\right)^\alpha & \mathrm{for} \quad  R<l_{ij}.\\
\end{array}\right.
\end{eqnarray}
Here, $\alpha$ and $R$ are positive real parameters. The radius $R$ determines a neighborhood around the current position $i$ of the walker, such that any other point part of this neighborhood can be visited with equal probability. Hence, these transitions are independent of the distance between the two sites. That is, if there are $S$ other sites inside the circle of radius $R$, the probability of jumping to any of these sites is constant. For the locations that are beyond the neighborhood of $i$, at distances greater than $R$, the transition probability decays as an inverse power-law with the distance and is proportional to $l_{ij}^{-\alpha}$ \cite{RiascosMateosPlos2017}. The parameter $R$ thus defines a neighborhood radius and $\alpha$ controls the probability that the walker performs long-range displacements. In particular, in the limit $\alpha\to \infty$ the dynamics becomes local, whereas in the case $\alpha\to 0$ the movement to any other point occurs with the same probability, namely, $w_{i\to j}^{(0)}(R)=(N-1)^{-1}$.
\\[2mm]
Having defined the random walker with long-range displacements between points, we consider $N=100$ locations distributed in the region $[0,1]\times [0,1]$ of $\mathbb{R}^2$ and represented in Fig. \ref{Fig_5}(a). The points are generated forming clusters with $10$, $20$, $30$ and $40$ points. Each cluster contains a possible resetting point (represented with a triangle) denoted by $\eta_1$, $\eta_2$, $\eta_3$, $\eta_4$. We use the value $R^\star\equiv \sqrt{\frac{\ln N}{\pi N}}$ as a reference length that, in the case of uniformly random distributed points, is the connectivity threshold of a random geometric graph \cite{Dall2002}. In our example, a random walker with transitions to nearest neighbors cannot reach all the points and is trapped in one of the clusters; however, the random walk strategy with long-range displacements and spatial information, given by Eq. (\ref{wijRalpha}), is ergodic for $0\leq \alpha <\infty$ and $R>0$.
\\[2mm]
Figure \ref{Fig_5}(b), displays the variations of the global mean first passage time $\mathcal{T}$ for the process generated by Eqs.   (\ref{wijRalpha})-(\ref{Omega_ij}), without resetting. We vary $\alpha$ for three values of $R$ in the transition matrix $w_{i\to j}^{(\alpha)}(R)$. For $R=1.5R^\star$, $\mathcal{T}$ depends moderately on $\alpha$, as far-away points are easily reached even through short-range steps, or large values of the exponent. In contrast, for the smaller value $R=0.5R^\star$, $\mathcal{T}$ depends sharply on $\alpha$, indicating that the network formed by connecting points distant by less than $R$ is practically disconnected, although the walker with $\alpha$ finite is capable of reaching any point. For $\alpha\gg 1$, the time $\mathcal{T}$ increases very rapidly, and a similar behavior is obtained for $R=R^\star$.
\\[2mm]
In  Figs. \ref{Fig_5}(c)-(d), $R$ is set to $R^\star$ and we study the global time $\mathcal{T}$ as a function of the total resetting probability $\beta$, for different values of $\alpha$ and $\mathcal{M}$. In Fig. \ref{Fig_5}(c), we consider resetting to the single point $\eta_4$ (that belongs to the largest cluster). Resetting does not impact substantially $\mathcal{T}$ for $\beta<10^{-3}$, independently of $\alpha$. For larger $\beta$, single-point resetting renders the exploration of the locations by the random walker rather inefficient, mostly for large $\alpha$, as much more time is needed to reach a site on average than without resetting. In Fig.  \ref{Fig_5}(d) we consider four resetting points $\eta_1,\ldots,\eta_4$ located in the four clusters, the resetting probability to each point being $\beta/4$. For $\alpha=1$ (bottom curve), the variations of $\mathcal{T}$ are small in the interval $0<\beta\leq 0.1$, like in the previous case with $\mathcal{M}=1$, which shows that these L\'evy flights constitute a very good exploration strategy, even in the absence of resetting. For $\alpha=5,10,15,20$, interestingly, the behavior becomes non-monotonic: multiple-point resetting reduces significantly the global time and a minimum is reached at a finite $\beta$. At this optimal point, resetting helps the walker to visit the four clusters, something that would require many more steps with one resetting point or in the absence of resetting. The results presented in Fig. \ref{Fig_5}(d) show that a broad distribution of resetting points can make exploration by a random walker with short steps nearly as efficient as a long-ranged L\'evy flight.
\subsection{Case $\mathcal{M}=N$, or the Google strategy}
\begin{figure*}[!t]
\begin{center}
\includegraphics*[width=1.0\textwidth]{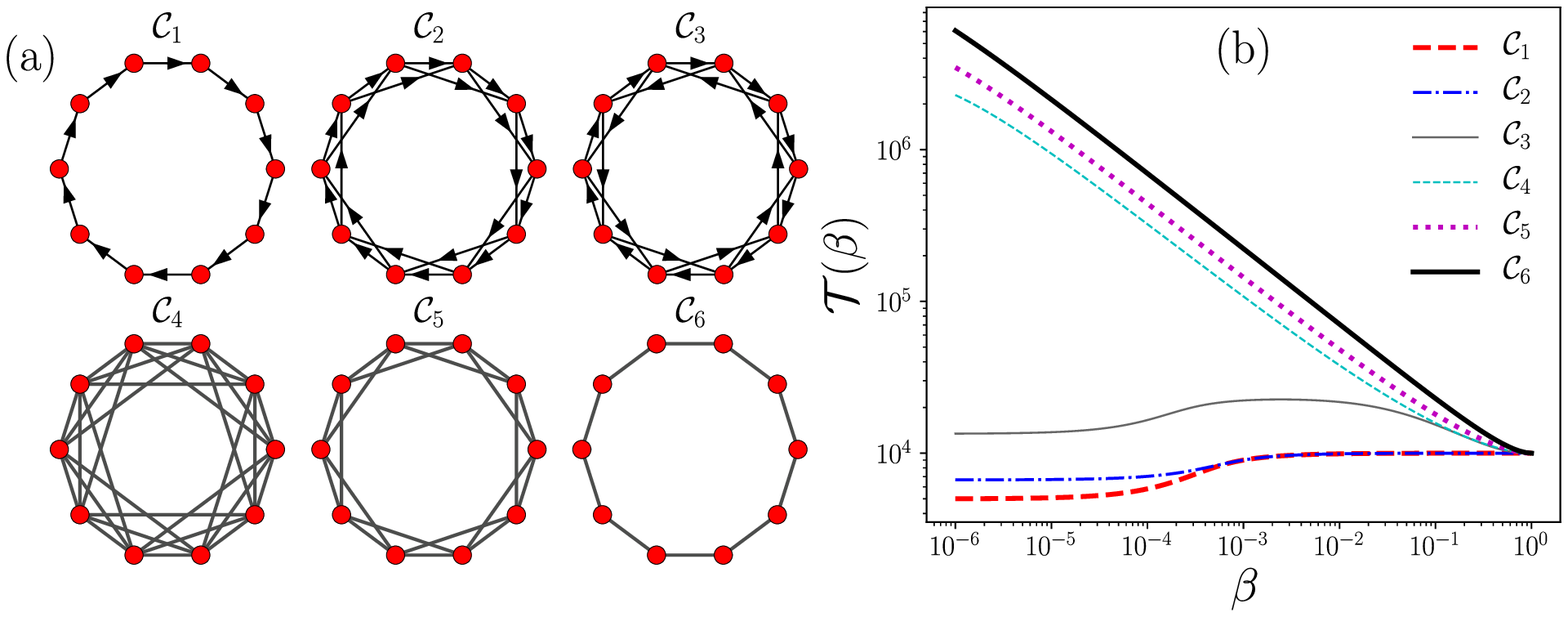}
\end{center}
\vspace{-5mm}
\caption{\label{Fig_6} Google random walks on different types of circulant networks. (a) Circulant networks with $N=10$ nodes, arrows represent the direction of each edge, and connections including both directions are represented with a line. (b) Global times $\mathcal{T}(\beta)$ in Eq. (\ref{tauglobal_google}) as a function of the total probability $\beta$ for circulant networks with $N=10^4$ nodes with the same topologies described in (a).}
\end{figure*}
The formalism introduced above with $\mathcal{M}$ resetting nodes can be applied to analyze the Google random walk strategy, that combines a local search to nearest-neighbor nodes with stochastic relocations to any of the $N$ nodes of the network with constant probability \cite{Brin1998,ShepelyanskyRevModPhys2015}. This case can be expressed in terms of resetting to multiple nodes by considering $\mathcal{M}=N$ and $a_1=a_2=\ldots=a_N$. If the total resetting probability is $\beta=\sum_{s=1}^N a_s$ then $a_l=\beta/N$ for $l=1,2,\ldots,N$. From Eq. (\ref{eigval_reset_ws}) and the definition (\ref{gamma_gendef}), the eigenvalues of the matrix that defines the Google strategy are   
\begin{equation}\label{eigvals_zeta_google}
\zeta^{\mathrm{Google}}_l=
\begin{cases}
1 \qquad &\mathrm{for}\qquad l=1,\\
(1-\beta)\lambda_l  &\mathrm{for}\qquad l=2,3,\ldots, N.
\end{cases}
\end{equation}
Hence the relation between the eigenvalues of the Google transition matrix and those of $\mathbf{W}$ is rather simple.
\\[2mm]
Here we explore the Google random walk strategy on circulant networks. These networks are such that both the adjacency matrix $\mathbf{A}$ and  $\mathbf{W}$  (with elements $w_{i\to j}=A_{ij}/{k_i}$) are circulant matrices \cite{VanMieghem2011}. For these structures $\mathbf{A}$ is not necessarily symmetric (the network can be directed) but $k_i=\sum_{l=1}^N A_{il}$ is constant or independent of $i$. On directed networks, $k_i$ represents the out-degree of $i$. Due to this regularity and the uniformity of resetting, the stationary distribution is constant and given by $1/N$. Therefore, for regular networks, the global MFPT defined by Eq. (\ref{Global_Tdef}) takes the form
\begin{align*}
&\mathcal{T}=\frac{1}{N^2}\sum_{i=1}^N\sum_{j=1}^N \langle T_{ij}\rangle\\
&=\frac{1}{N}\sum_{i=1}^N\sum_{j=1}^N\left[\delta_{ij}+\sum_{l=2}^{N}\frac{\langle j|\phi_{l}\rangle \langle \bar{\phi}_{l}|j\rangle - \langle i | \phi_{l}\rangle \langle \bar{\phi}_{l}|j\rangle}{1-(1-\beta)\lambda_{l}}\right].
\end{align*}
By using $\sum_{j=1}^N|j\rangle \langle j |=\mathbb{I}$ and from the orthonormalization between $\langle \bar{\phi}_{l}|$  and $|\phi_1\rangle$ in Eq. (\ref{phi_vec_conds}), $\sum_{j=1}^N \langle \bar{\phi}_{l}|j\rangle=0$ for $l=2,3,\ldots,N$, we obtain
\begin{equation}\label{tauglobal_google}
\mathcal{T}(\beta)=1+\sum_{l=2}^N\frac{1}{1-(1-\beta)\lambda_l}.
\end{equation}
The remaining task is to calculate the eigenvalues of $\mathbf{W}$. In a $N\times N$ circulant matrix $\mathbf{C}$ with elements $C_{ij}$, each column has real elements $c_0, c_1,\ldots,c_{n-1}$ ordered in such a way that $ c_0 $ describes the diagonal elements and $C_{ij}=c_{(i-j)\text{mod}\, N}$. The eigenvalues $\lambda_l$ of $\mathbf{C}$ are given by \cite{VanMieghem2011}
\begin{equation} \label{SpectCxi}
\lambda_l=\sum_{m=0}^{N-1} c_m e^{\text{i}\frac{2\pi}{N}(l-1)\, m}.
\end{equation}
With this general relation we can obtain analytically the eigenvalues $\lambda_l$ of $\mathbf{W}$ for a local random walker on a circulant network, directly from the coefficients $c_m$ that define $\mathbf{W}$ (see Ref. \cite{DirectedFractional_PRE2020} for a detailed discussion).
\\[2mm]
In Fig. \ref{Fig_6} we show $\mathcal{T}(\beta)$ on different circulant networks with $N$ nodes. The analyzed topologies are represented in Fig. \ref{Fig_6}(a): $\mathcal{C}_1$ represents a directed ring, with transition matrix defined by a single non-null element $c_1=1$; $\mathcal{C}_2$ defines a directed random walk with  $c_1=c_2=1/2$; $\mathcal{C}_3$ has $c_1=c_{N-2}=1/2$. In all these cases, the random walker moves on a directed network. Conversely, $\mathcal{C}_4$ (with non-null elements $c_1=c_2=c_3=c_{N-1}=c_{N-2}=c_{N-3}=1/6$), $\mathcal{C}_5$ ($c_1=c_2=c_{N-1}=c_{N-2}=1/4$), and the simple ring $\mathcal{C}_6$ ($c_1=c_{N-1}=1/2$) are undirected networks. In all these cases, the dynamics generated by $\mathbf{W}$ is ergodic and the respective eigenvalues are obtained from Eq. (\ref{SpectCxi}).
\\[2mm]
In Fig. \ref{Fig_6}(b) we represent the global time $\mathcal{T}(\beta)$ obtained from combining Eqs. (\ref{tauglobal_google}) and (\ref{SpectCxi}) for networks with the topologies described in Fig. \ref{Fig_6}(a) and $N=10^4$. We observe that relocating with probability $\beta/N$ the random walk to any node produces rather diverse and unexpected effects. For the directed cycles $\mathcal{C}_1$ and $\mathcal{C}_2$, the choice $\beta=0$ is optimal and $\mathcal{T}(\beta)$ slightly increases with $\beta$. For $\mathcal{C}_3$, the time $\mathcal{T}(\beta)$ varies little with $\beta$, too, but exhibits a local {\it maximum} and reaches its absolute minimum for $\beta\to 1$. In all the undirected networks $\mathcal{C}_4$, $\mathcal{C}_5$, $\mathcal{C}_6$, increasing $\beta$ greatly improves the capacity of the walker to explore the network and the optimum is reached for $\beta\to 1$.
\section{Conclusions}
We have studied the diffusion and first passage properties of discrete-time random walks on networks subject to resetting to more than one node. Multiple resetting is constructed by choosing a subset of $\mathcal{M}$ nodes in the network of size $N$ and by assigning a finite resetting probability to each node of this subset. At every single time step, the walker either jumps to a node according to a given transition probability matrix (for instance, to a nearest neighbor as in a standard random walk), or relocates to one of the resetting nodes with the corresponding resetting probability. In the limit where all the nodes act as resetting nodes ($\mathcal{M}=N$) and have the same resetting probability, the process is the so-called Google strategy. In this case, the global mean first passage time on circulant networks is given by the rather compact expression (\ref{tauglobal_google}), which is new to the best of our knowledge. The formalism developed here is quite general, though, and applicable to any kind of connected directed networks, or to any ergodic underlying process defined by transition probabilities between pairs of nodes, such as random walks on spatial networks, where the probabilities depend on the separation distance between the nodes. The basic quantities of interest can be calculated iteratively for $\mathcal{M}$ from the case $\mathcal{M}-1$, and computed ultimately in terms of the eigenvectors and eigenvalues of the random walk transition matrix in the absence of resetting.
\\[2mm]
Adding a resetting node to a single-node resetting process ($\mathcal{M}=2$) significantly decreases the mean first passage time at a target node located near the new resetting node, without strongly hindering the search for targets that are close to the original resetting node. This effect is illustrated by Fig. \ref{Fig_2} for the ring geometry. Therefore, the global MFPT, which quantifies the ability of the walker to explore a whole network, typically decreases with the number of resetting nodes, see, {\it e.g.}, Fig. \ref{Fig_4} for a comb graph. On directed graphs, however, the effects of resetting on the mean search time can be diverse and somehow unexpected, as shown for instance by the presence of a local {\it maximum} for the global MFPT with respect to the total resetting probability (Fig. \ref{Fig_6}).
\\[2mm]
Numerous studies since Refs. \cite{evans2011diffusion,Evans2011JPhysA} have shown that resetting typically makes a search process more efficient and that there often exists an optimal resetting rate for which the mean search time is minimal. In other cases, resetting is detrimental to search. A similar paradigm seems to emerge here for multiple-node resetting (varying the total resetting probability), with some important amendments, though. For instance, in Fig. \ref{Fig_5}, the MFPT is an increasing function of the resetting probability for a random walk under single-node resetting, whereas the same quantity becomes non-monotonous and exhibits a minimum if other resetting nodes are introduced. It would be interesting to seek a general principle able to predict when multiple-node resetting is beneficial to search, that would generalize the criterion exposed in Refs.  \cite{reuveni2016optimal,PalFP_PRL_2017,Belan_Restart_PRL2018} for standard resetting.
\section{Appendices}
\subsection{MFPTs for ergodic random walks}
\label{App_DeductionTij}
In this appendix we present the deduction of the mean first passage times for random walks on networks with stationary distribution $P_j^\infty$, $j=1,2,\ldots,N$. The results are general and can be implemented for the analysis of ergodic Markovian random walks. We apply an approach similar to the formalism presented in Refs. \cite{Hughes,NohRieger2004}. We start representing the occupation probability $P_{ij}(t)$ as \cite{Hughes}
\begin{equation}\label{EquF}
P_{ij}(t) = \delta_{t0} \delta_{ij} + \sum_{t'=0}^t   P_{jj}(t-t')  F_{ij}(t') \ ,
\end{equation}
where $F_{ij}(t^\prime)$ is the probability to reach the node $j$ for the first time after $t^\prime$ steps given the initial position $i$. By definition $F_{ij}(0)=0$, and $P_{jj}(t-t^\prime)$ is the probability to be located at the position $j$ again after $t-t^\prime$ steps. The first term in the right-hand side of Eq. (\ref{EquF}) enforces the initial condition.
\\[2mm]
By using the discrete Laplace transform $\tilde{f}(s) \equiv\sum_{t=0}^\infty e^{-st} f(t)$ in Eq. (\ref{EquF}), we have
\begin{equation}\label{LaplTransF}
\widetilde{F}_{ij} (s) = (\widetilde{P}_{ij}(s) - \delta_{ij}) / 
\widetilde{P}_{jj} (s) \ .
\end{equation}
In terms of $F_{ij}(t)$, the mean first passage time $\langle T_{ij}\rangle$ is given by \cite{Hughes}
\begin{equation}
\langle T_{ij}\rangle \equiv \sum_{t=0}^{\infty} t F_{ij} (t) = -\widetilde{F}'_{ij}(0).
\end{equation}
Using the moments $R^{(n)}_{ij}$ of the probability $P_{ij}(t)$ defined as
\begin{equation}
R^{(n)}_{ij}\equiv \sum_{t=0}^{\infty} t^n ~ \{P_{ij}(t)-P_j^\infty\},
\end{equation}
the expansion in series of $\widetilde{P}_{ij}(s)$ is
\begin{equation}
\widetilde{P}_{ij}(s) =P_j^\infty\frac{1}{(1-e^{-s})}
+ \sum_{n=0}^\infty (-1)^n R^{(n)}_{ij} \frac{s^n}{n!} \ .
\end{equation}
Introducing this result into Eq. (\ref{LaplTransF}), the MFPT is obtained
\begin{equation}\label{Tij}
\langle T_{ij} \rangle =\frac{1}{P_j^\infty}\left[R^{(0)}_{jj}-R^{(0)}_{ij}+\delta_{ij}\right] .
\end{equation}
\subsection{Deduction of Eq. (\ref{Integral_InfLine_M2})}
\label{Appendix_PartA}
We calculate the integral
\begin{equation}
\mathcal{I}_2=\frac{1}{2\pi}\int_{0}^{2\pi}\frac{\cos(\varphi x)}{(1-y\cos\varphi)(1-z\cos\varphi)}d\varphi,
\end{equation}
that appears in Eq. (\ref{PinfOnfRing_M2_integral}) for the stationary distribution of simple random walks on an infinite ring. Considering the partial fraction decomposition
\begin{multline}
\frac{1}{(1-y\cos\varphi)(1-z\cos\varphi)}\\
=\frac{1}{y-z}\left[\frac{y}{1-y\cos\varphi}-\frac{z}{1-z\cos\varphi}\right],
\end{multline}
one obtains
\begin{equation}\label{eq_I2asI1}
\mathcal{I}_2=\frac{y\,\mathcal{I}_1(y,x)-z\,\mathcal{I}_1(z,x)}{y-z},
\end{equation}
where we have used the identity \cite{ResetNetworks_PRE2020}
\begin{align}\nonumber
\mathcal{I}_1(b,x)&=\frac{1}{2\pi}\int_0^{2\pi} \frac{\cos(x\varphi)}{1-b\cos(\varphi)}d\varphi\\
&=\frac{\left(\frac{1+\sqrt{1-b^2}}{b}\right)^{-x}}{\sqrt{1-b^2}}.
\end{align}
Therefore, Eq. (\ref{eq_I2asI1}) gives
\begin{equation}
\mathcal{I}_2=\frac{y}{y-z}\frac{\left(\frac{1+\sqrt{1-y^2}}{y}\right)^{-x}}{\sqrt{1-y^2}}-
\frac{z}{y-z}\frac{\left(\frac{1+\sqrt{1-z^2}}{z}\right)^{-x}}{\sqrt{1-z^2}}.
\end{equation}
\subsection{L\'evy flights with reset on an infinite ring}
\label{Appendix_PartB}
In this part, we specify the stationary distribution of  L\'evy flights on an infinite ring with resetting probability $\gamma$ to the single node $r$. Considering the limit $N\to \infty$ for the stationary distribution in Eq. (\ref{Pinf_levyfinite_ring}), we obtain ($0<\alpha<1$)
\begin{equation}\label{Pinf_IntegralLevyM1}
P_{j}^{\infty}(r;\gamma)=
\frac{\gamma}{2\pi}\int_{0}^{2\pi}\frac{\cos(d_{j r}\varphi)d\varphi}{1-(1-\gamma)[1-\frac{(2-2\cos\varphi)^\alpha}{k^{(\alpha)}}]}.
\end{equation}
The limit $\alpha\to 1$ recovers the result discussed in Ref.  \cite{ResetNetworks_PRE2020} for nearest-neighbor random walks on an infinite ring. L\'evy flights in Eq. (\ref{Pinf_IntegralLevyM1}) correspond to $\alpha<1$, where $k^{(\alpha)}$ is the fractional degree defined by Eq. (\ref{degreeGLring}). For an infinite ring, it takes the form \cite{RiascosMateosFD2015}
\begin{equation}
k^{(\alpha)}=-\frac{\Gamma(-\alpha)\Gamma(1+2\alpha)}{\pi\Gamma(1+\alpha)}\sin(\pi \alpha).
\end{equation}
We have
\begin{align}\nonumber
P_{j}^{\infty}(r;\gamma)&=\frac{\gamma}{2\pi}\int_{0}^{2\pi}\frac{\cos(d_{j r}\varphi)d\varphi}{1-(1-\gamma)[1-\frac{2^{2\alpha}}{k^{(\alpha)}}\sin(\varphi/2)^{2\alpha}]}\\ \nonumber
&
=\frac{\gamma}{\pi}\int_{0}^{\pi}\frac{\cos(2d_{j r}\theta)}{1-(1-\gamma)[1-\frac{2^{2\alpha}}{k^{(\alpha)}}(\sin\theta)^{2\alpha}]}d\theta\\ \nonumber
&
=\frac{1}{\pi}\int_{0}^{\pi}\frac{\cos(2d_{j r}\theta)}{1+\frac{(1-\gamma)}{\gamma}\frac{2^{2\alpha}}{k^{(\alpha)}}(\sin\theta)^{2\alpha}}d\theta\\
&
=\frac{1}{\pi}\int_{0}^{\pi}\frac{\cos(2d_{j r}\theta)}{1+D_\alpha(\sin\theta)^{2\alpha}}d\theta,
\label{Pinf_IntegralLevy_M1Dalpha}
\end{align}
where $D_\alpha=\frac{(1-\gamma)}{\gamma}\frac{2^{2\alpha}}{k^{(\alpha)}}$. However, using the series expansion of the denominator in Eq. (\ref{Pinf_IntegralLevy_M1Dalpha})
\begin{equation}\label{PinfLevy_series}
P_{j}^{\infty}(r;\gamma)=\frac{1}{\pi}\sum_{n=0}^\infty(-D_\alpha)^n\int_{0}^{\pi}(\sin\theta)^{2\alpha n}\cos(2d_{j r}\theta)d\theta.
\end{equation}
Let us now consider the asymptotic limit $d_{j r}\gg 1$. We use the identities
\begin{align}\nonumber
&\int_{0}^{\pi}(\sin\theta)^{2\alpha n}\cos(2 x \theta)d\theta\\\nonumber
&=\frac{2^{-2\alpha n} \pi\cos(\pi x)\Gamma(1+2\alpha n)}{\Gamma(1+\alpha n-x)\Gamma(1+\alpha n+x)}\\
&=-2^{-2\alpha n} \sin(\pi\alpha n)\Gamma(1+2\alpha n)\frac{\Gamma(x-\alpha n)}{\Gamma(x+1+\alpha n)}.
\end{align}
For $x\gg 1$, $\Gamma(x+m)\sim \Gamma(x)x^m$, therefore
\begin{equation}\label{GammaRel_xlarge}
\frac{\Gamma(x-\alpha n)}{\Gamma(x+1+\alpha n)}\sim \frac{1}{x^{1+2\alpha n}}.
\end{equation}
By combining Eqs. (\ref{PinfLevy_series})-(\ref{GammaRel_xlarge}), we obtain for $d_{jr}\gg 1$
\begin{equation}
P_{j}^{\infty}(r;\gamma)\sim\sum_{n=1}^\infty\frac{(-1)^{n+1}D_\alpha^n  \sin(\pi\alpha n)\Gamma(1+2\alpha n)}{\pi\,2^{2\alpha n}\,d_{jr}^{1+2\alpha n}},
\end{equation}
whose leading term is given by
\begin{align}\nonumber
P_{j}^{\infty}(r;\gamma)&\sim\frac{D_\alpha\sin(\pi\alpha)\Gamma(1+2\alpha)}{\pi 2^{2\alpha}}\frac{1}{d_{jr}^{1+2\alpha}}+\ldots\\ \nonumber
&\sim\frac{(1-\gamma)\sin(\pi\alpha)\Gamma(1+2\alpha)}{\pi \gamma k^{(\alpha)}}\frac{1}{d_{jr}^{1+2\alpha}}+\ldots\\
&\sim-\left(\frac{1-\gamma}{\gamma}\right)\frac{\Gamma(1+\alpha)}{\Gamma(-\alpha)}\frac{1}{d_{jr}^{1+2\alpha}}+\ldots.
\end{align}
This result serves as a guide for understanding the asymptotic dependence of the stationary distribution $P_{j}^{\infty}(r;\gamma)$ with the distance $d_{jr}$. The result in integral form in Eq. (\ref{Pinf_IntegralLevy_M1Dalpha}) is exact; however, the relation obtained from the series expansion in Eq. (\ref{PinfLevy_series}) is conditioned to its convergence. A different approach for the analysis of the effect of resetting on L\'evy flights using fractional dynamics on an infinite continuous line establishes the same asymptotic relation $P_{j}^{\infty}(r;\gamma)\sim\frac{1}{d_{jr}^{1+2\alpha}}$ as above, but for $1/2\leq \alpha<1$ (see Ref. \cite{kusmierz2015optimal} for a detailed discussion).
\\[2mm]
In addition, applying the same approach to the analysis of the MFPT in Eq. (\ref{MFPT_levy_ring_finite}) for L\'evy flights in the limit $N\to \infty$, we have 
\begin{multline}\left\langle T_{ij}(i;\gamma)\right\rangle=\frac{1}{\gamma}+\frac{1}{P_j^\infty(i;\gamma)}\times\\
\left[\delta_{ij}+\frac{1}{2\pi}\int_0^{2\pi}\frac{d\varphi}{1-(1-\gamma)\left[1-\frac{2^{2\alpha}}{k^{(\alpha)}}\sin(\varphi/2)^{2\alpha}\right]}\right]. 
\end{multline}
However
\begin{align*}&\frac{1}{2\pi}\int_0^{2\pi}\frac{d\varphi}{1-(1-\gamma)\left[1-\frac{2^{2\alpha}}{k^{(\alpha)}}\sin(\varphi/2)^{2\alpha}\right]}\\
&=\frac{1}{\pi}\int_0^{\pi}\frac{d\theta}{1-(1-\gamma)\left[1-\frac{2^{2\alpha}}{k^{(\alpha)}}\sin(\theta)^{2\alpha}\right]}\\
&=\frac{1}{\gamma\pi}\int_0^{\pi}\frac{d\theta}{1+\frac{(1-\gamma)}{\gamma}\frac{2^{2\alpha}}{k^{(\alpha)}}\sin(\theta)^{2\alpha}}\\
&=\frac{1}{\gamma\pi}\int_0^{\pi}\frac{d\theta}{1+D_\alpha\sin(\theta)^{2\alpha}}=\frac{1}{\gamma}\mathcal{G}(\gamma,\alpha),
\end{align*}
where we defined $\mathcal{G}(\gamma,\alpha)$ that depends on $\gamma$, $\alpha$ but is independent of the distance between $i$ and $j$. Therefore
\begin{equation}\left\langle T_{ij}(i;\gamma)\right\rangle=\frac{1}{\gamma}+\frac{1}{P_j^\infty(i;\gamma)}\left[\delta_{ij}+\frac{\mathcal{G}(\gamma,\alpha)}{\gamma}\right]. 
\end{equation}
In this way, for  $0<\gamma<1$, $1/2\leq\alpha<1$
\begin{equation}
\left\langle T_{ij}(i;\gamma)\right\rangle\sim d_{ij}^{1+2\alpha},\qquad d_{ij}\gg 1.
\end{equation}
A relation that agrees with the result reported in Ref. \cite{kusmierz2015optimal}.
\section*{Acknowledgments}
The authors acknowledge support from Ciencia de Frontera 2019 (CONACYT), project ``Sistemas complejos estoc\'asticos: Agentes m\'oviles, difusi\'on de part\'iculas, y din\'amica de espines'' (Grant No. 131, key 10872).


%

\end{document}